\documentclass[twocolumn,showpacs,pra,amsmath,amssymb,eqsecnum]{revtex4}

\usepackage[T1]{fontenc}
\usepackage[latin9]{inputenc}

\usepackage{esint}
\usepackage{epsfig}
\usepackage{amsmath}
\usepackage{latexsym}

\usepackage{amssymb,amsmath}

\usepackage{graphicx}
\usepackage{dcolumn}
\usepackage{bm }
\usepackage{epstopdf}
\begin{document}

\def\be{\begin{equation}}
\def\ee{\end{equation}}

\title{Eigenstates of the full Maxwell equations for a two-constituent\\
composite medium and their application
to a calculation of the local electric field\\of a time dependent point electric dipole
in a flat-slabs microstructure}

\author{Asaf Farhi}
 \email{asaffarhi@post.tau.ac.il}
\author{David J. Bergman}
\email{bergman@post.tau.ac.il}

\affiliation{
Raymond and Beverly Sackler School of Physics and Astronomy,
Faculty of Exact Sciences,
Tel Aviv University, IL-69978 Tel Aviv, Israel
}

\date{\today}

\begin{abstract}

An exact calculation of the local electric field ${\bf E}({\bf r})$ is described for the case of a
time dependent point electric dipole ${\bf p}e^{-i\omega t}$ in the top layer of an  $\epsilon_2$, $\epsilon_1$, $\epsilon_2$ three parallel slabs composite structure, where the  $\epsilon_1$ layer
has a finite thickness  $2d$ but the $\epsilon_2$ layers are infinitely thick.
For this purpose we first calculate all the eigenstates of the full Maxwell equations for the case
where $\mu=1$ everywhere in the system. The eigenvalues appear as special,
non-physical values of $\epsilon_1$ when $\epsilon_2$ is given. These eigenstates
are then used to develop an exact
expansion for the physical values of ${\bf E}({\bf r})$ in the system characterized by
physical values of $\epsilon_1(\omega)$ and $\epsilon_2(\omega)$. Results are
compared with those of a previous calculation of the local field of a time dependent point
charge in the quasi-static regime. Numerical results are shown for the local electric field
in practically important configurations where attaining an optical image with
sub-wavelength resolution has practical significance.

\end{abstract}



\maketitle




\section{INTRODUCTION}
\label{introduction}

In order to have a physical electromagnetic (EM) field in some system volume it is usually
necessary to have either a field incident from outside of the system volume
or a non-vanishing charge density and current density inside the system. However,
when the material in the system has certain special values of its material parameters,
a field can arise in the system spontaneously. Such a state is an EM eigenstate
and the special parameters are the appropriate eigenvalues.
While such eigenstates can never be realized in a passive physical system, because the
necessary values of its material parameters are unachievable in a real material, these states
are often useful in particular circumstances. Thus, a real material can have parameters that approach
some of the eigenvalues, in which case the EM response of such a material
can become anomalously strong. Furthermore, the EM field of a real physical system can
 be expanded in a series of the eigenstates, leading to an alternative approach
to the calculation of that field and its consequences. Such an approach was used in the
past to describe the scattering of EM radiation by a collection of spheres \cite{BergStroudPRB80}.
Such an approach was also applied, in the past, to calculate the macroscopic response
of a collection of spheres in the quasi-static regime, i.e., the macroscopic electric permittivity
$\epsilon_e$ of such a material \cite{BergPRB79,BergJPC79}. More recently, such an approach was used to
compute the local electric field in a special structure, known as the Veselago Lens \cite{Veselago}, where
it had been claimed that an EM image was achievable with unrestricted resolution \cite{Pendry}.
By exploiting an expansion of the local electric field in the exact quasi-static eigenstates for the case of a point charge, a much
more detailed analysis of this system became possible \cite{BergPRA2014,FarhiBergVeselagoPRA2014}. In previous discussions the asymptotic expression for the potential at the interface between the lens and the medium when $\epsilon_1=-\epsilon_2$, both real, was shown to diverge \cite{MiltonPRB1994,MiltonProcRoySoc2005,MiltonProcRoySoc2006}. In Refs. \cite{BergPRA2014,FarhiBergVeselagoPRA2014} an exact expression for the potential (in all space) in the form of a 1D integral was derived for general complex permittivity values and it was shown that the imaging, in terms of both intensity and resolution, is optimal at the interface.
In recent works a 2D setup of a coated cylinder with an external line source was analyzed using the full Maxwell equations and the asymptotic expression for the electric field at the interface when $\epsilon_1=-\epsilon_2$, both real, was shown to diverge \cite{MiltonProcRoySoc2006,KettunenArxiv2014}. 
In this article we attempt to extend the approach of Refs. \cite{BergPRA2014,FarhiBergVeselagoPRA2014} to expand the electric field (in all space) in the exact eigenstates of the full Maxwell equations for complex permittivity values and a general 3D current distribution. This will be used to extend the discussion of a Veselago Lens
to the non-quasi-static regime. The formalism enables to calculate the electric field also for current sources in a simple manner, avoiding the complex calculation of the scattering of the electric field of these sources.  

The general theory for this is developed in Section \ref{eigenstates}.
In Section \ref{FlatSlabs} the eigenstates are calculated in closed form for the special structure
of a flat slab, which is also the structure of the Veselago Lens. In Section \ref{section:calculation}
these eigenstates are used to expand the local EM field produced in such a lens by an
oscillating point electric dipole source directed parallel and perpendicular to the slab. Section \ref{discussion} includes a summary of our
main results and a discussion of possible future extensions of the approach developed here.

\section{Theory of the eigenstates of Maxwell's equations in a two-constituent composite
medium where {\large $\mu=1$}}

\label{eigenstates}

We assume that all physical quantities are monochromatic functions of time, namely that
they are proportional to $e^{-i\omega t}$.
We confine ourselves to the case where $\mu=1$ everywhere,
but the position dependent electric permittivity $\epsilon({\bf r})$ has two different
values corresponding to a two-constituent composite medium:
\be
\epsilon({\bf r})=\epsilon_1\theta_1({\bf r})+\epsilon_2\theta_2({\bf r}),
\label{epsilonr}
\ee
where $\theta_i({\bf r})$, $i=1,2$ is a step function equal to 1 when
${\bf r}$ is inside the $\epsilon_i$ constituent and equal to 0 elsewhere.
Note that $\epsilon_i$ is usually complex and includes any electrical conductivity
that the constituents may have.
Assuming that all the EM fields are monochromatic and a general current distribution, Maxwell's equations
become, in Gaussian units,
\begin{eqnarray}
\nabla\cdot(\epsilon{\bf E})=0,\;\nabla\times{\bf E}=\frac{i\omega}{c}{\bf H},\; \nonumber \\
 \nabla\cdot{\bf H}=0,\; \nabla\times{\bf H}=-\frac{i\omega}{c}\epsilon{\bf E}+\frac{4\pi}{c}\mathbf{J}.
\label{Maxwell}
\end{eqnarray}
From these we can obtain the following equation for the local electric field
${\bf E}({\bf r})$:
\begin{eqnarray}
-\nabla\times(\nabla\times{\bf E})+k_2^2{\bf E}=uk_2^2\theta_1{\bf E}-\frac{4\pi i\omega}{c^{2}}\mathbf{J}, \label{EdiffEq}\\
u \equiv 1-\frac{\epsilon_1}{\epsilon_2},\;\;\;\;k_2^2\equiv\epsilon_2\frac{\omega^2}{c^2}.
\label{uk2}
\end{eqnarray}

The last differential equation can be transformed into an integral equation by
using a tensor Green function $G_{\alpha\beta}({\bf r},{\bf r}',k_2)$, defined by
\be
-\nabla\times(\nabla\times\stackrel{\leftrightarrow}{G})+k_2^2\stackrel{\leftrightarrow}{G}=
k_2^2\openone\delta^3({\bf r}-{\bf r}')
\label{GreenEq}
\ee
and by appropriate outgoing boundary conditions at large distances $|{\bf r}-{\bf r}'|$.
Noting that $G_{\alpha\beta}({\bf r},{\bf r}',k)$ will depend on those position vectors
only through their difference ${\bf R}\equiv{\bf r}-{\bf r}'$, we
first apply a spatial Fourier transformation to this equation. This results in a linear algebraic
equation for the Fourier transform of $G_{\alpha\beta}({\bf r}-{\bf r}',k)$ which is easily solved,
leading to the following expression for that Fourier transform:
\be
G_{\alpha\beta}({\bf q},k)=\frac{q_\alpha q_\beta-k^2\delta_{\alpha\beta}}{{ q}^2-k^2}.
\label{FTgreen}
\ee
The inverse Fourier transform of this, with the boundary condition of an outgoing or evanescent
wave at large distances, is found by first integrating over the direction of the three-dimensional vector
{\bf q}, leading to the
remaining integral over the magnitude of {\bf q} ($q\equiv|{\bf q}|$):
\begin{eqnarray}
G_{\alpha\beta}({\bf R},k)&=&(k^2\delta_{\alpha\beta}+\nabla_\alpha\nabla_\beta)
  \frac{i}{(2\pi)^2|{\bf R}|}\int q\,dq\frac{e^{iqR}}{q^2-k^2}\label{GsphericalA}\nonumber \\
&=-&(k^2\delta_{\alpha\beta}+\nabla_\alpha\nabla_\beta)\frac{e^{ikR}}{4\pi R}.
\label{Gspherical}
\end{eqnarray}
The last integration here was carried out by adding to the real axis of $q$ an infinite radius semi-circle
in the upper complex plane of $q$ and then using Cauchy's theorem to evaluate the integral
over the resulting closed contour.
This closed form expression for $\stackrel{\leftrightarrow}{G}({\bf r}-{\bf r}',k)$ was obtained many years ago
in Ref. \cite{BergStroudPRB80}.

Using $\stackrel{\leftrightarrow}{G}({\bf r}-{\bf r}',k_2)$ we can now ``solve''
Eq.\ (\ref{EdiffEq}) by treating its rhs as if it were known. In this way we get the following
integral equation for the local electric field ${\bf E}({\bf r})$:
\begin{eqnarray}
{\bf E}&=&{\bf E_0}+u\hat\Gamma{\bf E},
\label{IntegEq}\\
\hat\Gamma{\bf E}&\equiv&\int dV'\theta_1({\bf r}')\stackrel{\leftrightarrow}{G}({\bf r}-{\bf r}',k_2)
 \cdot{\bf E}({\bf r}'),
\label{GammaDef}
\end{eqnarray}
where $\bf E_0$ is the the electric field generated by the external sources $\mathbf{J}\left(\mathbf{r}\right)$ in a uniform $\epsilon_2$ medium.

The scalar product of two vector fields ${\bf F}({\bf r})$, ${\bf E}({\bf r})$ is now defined by
\be
\langle{\bf F}|{\bf E}\rangle\equiv\int dV\theta_1({\bf r}){\bf F}^*({\bf r})\cdot{\bf E}({\bf r}).
\label{ScalarProdDef}
\ee
Under this definition $\hat\Gamma$ is a symmetric operator, as defined in Appendix \ref{biorthogonal},
because $G_{\alpha\beta}({\bf R},k)=G_{\beta\alpha}(-{\bf R},k)$, but it is non-Hermitian 
because $\stackrel{\leftrightarrow}{G}({\bf r}-{\bf r}',k_2)$ is complex valued.
Thus the left eigenstates of $\hat\Gamma$ $,\left\langle \tilde{{\bf E}}_{n}\right|$ are just the dual states of 
 its right eigenstates and  the left and right eigenvalues are the same:
\begin{eqnarray}
\lefteqn{
\langle\tilde{\bf E}_n|\hat\Gamma|{\bf r}\rangle\equiv\int dV' \theta_1({\bf r}'){\bf E}_n({\bf r}')\cdot
 \stackrel{\leftrightarrow}{G}({\bf r}'-{\bf r},k)}\nonumber\\
&=&\int dV' \theta_1({\bf r}')
 \stackrel{\leftrightarrow}{G}({\bf r}-{\bf r}',k)\cdot{\bf E}_n({\bf r}')=
  \langle{\bf r}|\hat\Gamma|{\bf E}_n\rangle\nonumber\\ 
&&\hspace{-5 true mm}\Longrightarrow s_n|{\bf E}_n\rangle=\hat\Gamma|{\bf E}_n\rangle,\;\;\;\;
\label{EigenstateEq}
s_n\langle\tilde{\bf E}_n|=\langle\tilde{\bf E}_n|\hat\Gamma,
\label{ConjEigenstates}
\end{eqnarray}
where $\langle\tilde{\bf E}_n|{\bf r}\rangle=\langle{\bf r}|\tilde{\bf E}_n\rangle^*\equiv
 \langle{\bf r}|{\bf E}_n\rangle\equiv{\bf E}_n({\bf r})$.

Because $\hat\Gamma$ is  a symmetric operator it therefore has the following property for any two states
$|{\bf E}\rangle$ and $|{\bf F}\rangle$
$$
\langle\tilde{\bf F}|\hat\Gamma|{\bf E}\rangle=\langle\tilde{\bf E}|\hat\Gamma|{\bf F}\rangle.
$$
From this it is now easy to show that the eigenstates and their duals satisfy
\be
\langle\tilde{\bf E}_n|{\bf E}_m\rangle=0
\label{EigScalarProd}
\ee
if $s_n\neq s_m$.

The scalar product of a left eigenstate and a right eigenstate of $\hat\Gamma$ can be
written as follows:
\be
\langle\tilde{\bf F}_n|{\bf E}_m\rangle=\int dV\theta_1({\bf r}){\bf F}_n({\bf r})\cdot{\bf E}_m({\bf r}).
\label{ScalarProdLeftRightEigen}
\ee
This differs from Eq.\ (\ref{ScalarProdDef}) because the dual eigenfunction
$\langle{\bf r}|\tilde{\bf F}_n\rangle$ is not equal to the eigenfunction
$\langle{\bf r}|{\bf F}_n\rangle$ but rather to its complex conjugate
$\langle{\bf r}|{\bf F}_n\rangle^*$.
Clearly, in the general case where these eigenfunctions are complex valued this scalar product  is not
assured to be real or positive and could vanish
even when the states $|{\bf E}_m\rangle$ and $|{\bf F}_n\rangle$ are the same, because  the integrand
is $\theta_1({\bf r})[{\bf E}_n({\bf r})]^2$ and not $\theta_1({\bf r})|{\bf E}_n({\bf r})|^2$. Thus, the question of normalizability
of the eigenstates must be investigated for each of them separately. We will nevertheless
assume that they are normalizable in our case and that they therefore form a complete set.
Thus, from the pair of equations (\ref{EigenstateEq}) we conclude that
the unit operator can be expanded in terms of those states and their duals
$|\tilde{\bf E}_n\rangle$ as
\be
	\openone=\sum_{n}\frac{|\mathbf{E}_{n}\rangle\langle{\bf \tilde{E}}_{n}|}{\langle\tilde{{\bf E}}_{n}|{\bf E}_{n}\rangle}.
\label{UnitOp}
\ee

We can now write the following formal solution of Eq.\ (\ref{IntegEq}):
\begin{eqnarray}
|{\bf E}\rangle&=&\frac{1}{1-u\hat\Gamma}|{\bf E}_0\rangle=|{\bf E}_0\rangle+
 \frac{\hat\Gamma}{s-\hat\Gamma}|{\bf E}_0\rangle,\\
 s&\equiv&\frac{1}{u}\equiv \frac{\epsilon_{2}}{\epsilon_{2}-\epsilon_{1}} ,
\end{eqnarray}
and insert the unit operator of Eq.\ (\ref{UnitOp}) to obtain
\be
|{\bf E}\rangle-|{\bf E}_0\rangle=\sum_n\frac{s_n}{s-s_n}|{\bf E}_n\rangle
\frac{\langle\tilde{\bf E}_n|{\bf E}_0\rangle}{\langle\tilde{\bf E}_n|{\bf E}_n\rangle},
\label{Eexpansion}
\ee
where $s_{n}\equiv\epsilon_{2}/\left(\epsilon_{2}-\epsilon_{1,n}\right)$ and $\epsilon_{1,n}$ is the eigenvalue which corresponds to $s_n.$

It is now useful to recall that the eigenstates of $\hat\Gamma$ fall into two classes
\cite{BergStroudPRB80}:
\begin{enumerate}

\item {\em Longitudinal eigenstates} ${\bf E}({\bf r})=\nabla\phi({\bf r})$. For these
states the eigenvalue will always be $s=1$ and ${\bf E}({\bf r})$ must vanish outside the
$\epsilon_1$ volume. Inside that volume ${\phi}({\bf r})$ is almost arbitrary and the
differential equation (\ref{EdiffEq})  is satisfied in a trivial fashion. The only restriction
on ${\phi}({\bf r})$ is due to the fact that the tangential component of ${\bf E}({\bf r})$
must be continuous at the $\epsilon_1$, $\epsilon_2$ interface. Since  its
tangential component must vanish there, therefore ${\phi}({\bf r})$ must be constant over every connected piece of that interface. Obviously, the magnetic field ${\bf H}({\bf r})$ vanishes everywhere for
these states.

\item {\em All the eigenstates for which $s\neq 1$ (these are transverse fields)}. From Eq.\
(\ref{EdiffEq}) it follows that $\nabla\cdot{\bf E}=0$ inside both the $\epsilon_1$ and
the $\epsilon_2$ regions, though not at their interface. These states must obey Eq.\ (\ref{EdiffEq})
in a nontrivial fashion.

\end{enumerate}

Although the Class 1 eigenstates are difficult to catalog, since they have a degenerate
eigenvalue, it turns out that they are not needed for expanding any physical solutions of
Eq.\ (\ref{EdiffEq}). That is because they are orthogonal to any solution ${\bf E}({\bf r})$
of Maxwell's equations. To see this, we denote by ${\bf E}_1({\bf r})\equiv\nabla\phi_1$
any longitudinal eigenstate and write
$$
\langle\tilde{\bf E}_1|{\bf E}\rangle=\int dV\theta_1 {\bf E}_1\cdot{\bf E}=
 \int_{V_1}dV\left[\nabla\cdot(\phi_1{\bf E})-\phi_1\nabla\cdot{\bf E}\right],
$$
where $V_1$ is the $\epsilon_1$ subvolume.
The second term under the last integral vanishes because the field ${\bf E}({\bf r})$
is a transverse field inside $V_1$ as long as $\epsilon_1\neq 0$---see Eq.\ (\ref{Maxwell}).
(Note that in a real physical material $\epsilon_1$ can never vanish: It must always
have a nonzero imaginary part which represents dissipation.)
The first term can be transformed into a surface integral over the
$\epsilon_1$, $\epsilon_2$ interface, where $\phi_1$ is a constant, denoted by
$\phi_{1i}$, over every connected portion
of that interface. Transforming the surface integral back to a sum of volume integrals
over the different connected pieces $V_i$ of $V_1$, where $\phi_1$ is replaced by
$\phi_{1i}$ which is constant over any connected volume piece $V_i$,
each of those integrals can be written as
$$
\phi_{1i}\int_{V_i}dV\nabla\cdot{\bf E}=0.
$$
We have thus shown that
$\langle\tilde{\bf E}_1|{\bf E}\rangle=0$ for $u\neq 1$.

The physical significance of the Class 2 eigenstates is that at special values of 
$\epsilon_1/\epsilon_2$ (the eigenvalues) a wave can arise in the system
spontaneously, i.e., without the presence of an incident wave or any source of radiation.
Since the fields are periodic in time, the local energy density must be constant when
averaged over one period. However, if the eigenfunction is an outgoing propagating wave at
large distances
then it constantly radiates energy out to infinity. In order to preserve the local energy density
the system must therefore create energy. For this to happen then at least one of the
two constituent permittivities must have an imaginary part with the ``wrong sign''.
Thus, if $\epsilon_2$, which is where the outgoing wave must propagate, has a physically
admissible value with an imaginary part that has the right sign, then $\epsilon_1$ will
have to have an imaginary part with the wrong sign. This means, of course, that the
system can never actually be at a resonance, but can only approach it if the
magnitude of the right signed physical Im$\,\epsilon_2$, as well as the
magnitude of the wrong signed eigenvalue Im$\,\epsilon_1$,
are small. On the other hand, if the eigenfunction decays exponentially at large
distances, and thus no energy is radiated, then the special values of 
$\epsilon_1/\epsilon_2$ and $s\equiv\epsilon_2/(\epsilon_2-\epsilon_1)$ can be real.

If all the eigenvalues $s_n$ are non-degenerate then the above analysis often suffices to fix the states
$|{\bf E}_n\rangle$ as a basis of Hilbert space, subject to their normalizability.
However, if the system has some symmetries, which are represented by Hermitian or
unitary operators $\hat P_i$ that commute
with $\hat\Gamma$, then that complicates the situation: We usually try to characterize the
eigenstates of $\hat\Gamma$ by requiring them to also be eigenstates of those symmetry operators.
However, since the operator $\hat\Gamma$ is symmetric we defined $\langle\tilde{\bf E}_n|{\bf r}\rangle\equiv \langle{\bf r}|{\bf E}_n\rangle,$ whereas the requirement for the eigenstates of Hermitian operators is that \mbox{$\langle\psi|{\bf r}\rangle=\langle{\bf r}|\psi\rangle^{*}.$}
An example of a symmetry generator is the infinitesimal spatial translation operator $-i\nabla,$ where
$$
-i\nabla e^{i{\bf q}\cdot{\bf r}}={\bf q} e^{i{\bf q}\cdot{\bf r}},$$
which is relevant for any microstructure that has translational symmetry along certain directions.
Clearly, the complex conjugate of any such eigenfunction will have the different eigenvalue
$-{\bf q}$. This problem can be overcome as follows: In the subspace of the complex conjugates
of all the right eigenfunctions $\langle{\bf r}|{\bf F}_n\rangle$ of $\hat\Gamma$ with the same
eigenvalue $s_n$ as $\langle{\bf r}|{\bf E}_n\rangle$ we choose one such that
$\langle{\bf r}|{\bf F}_n\rangle^*$ is a right eigenfunction of $-i\nabla$ with the same
eigenvalue as $\langle{\bf r}|{\bf E}_n\rangle$. This is done in Section \ref{FlatSlabs}
below for the particular case of a flat slabs microstructure.

\section{Eigenstates of a flat slabs microstructure}
\label{FlatSlabs}

\begin{figure}[t]
   \begin{center}
   \begin{tabular}{c}
   \includegraphics[height=5cm]{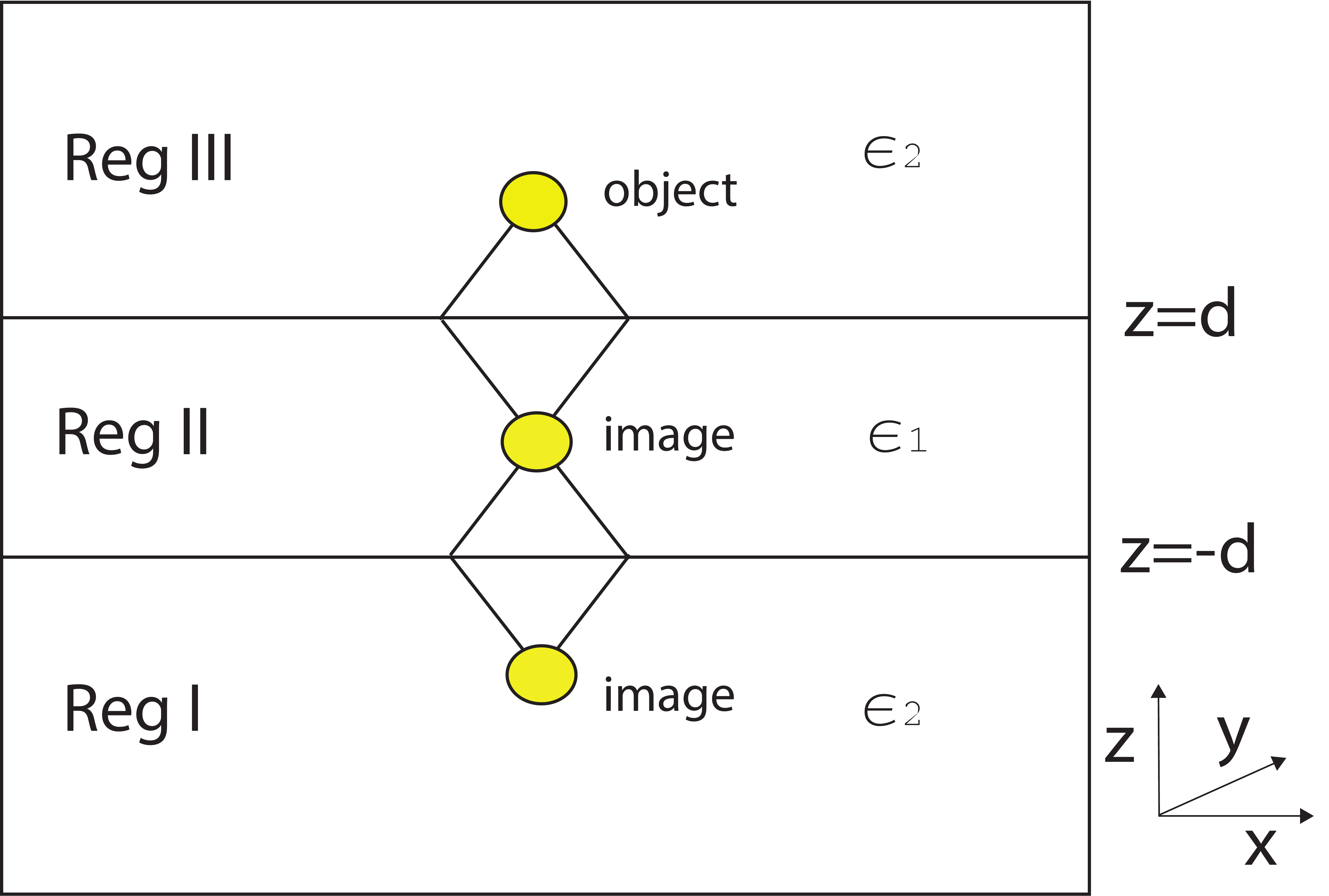}
   \end{tabular}
   \end{center}
   \caption[]
   { \label{fig:capacitor} 
An illustration of the setup of an $\epsilon_1$ slab in an $\epsilon_2$ medium. The object and images according to geometrical optics are represented by circles.}
   \end{figure}

Consider a medium with electric permittivity $\epsilon_2$ in which there is a flat slab,
of thickness $2d$,
with a different electric permittivity $\epsilon_1$ (see Fig.\ \ref{fig:capacitor}).
The magnetic permeabilty is
everywhere equal to 1, as in the vacuum [here and throughout this article we use
Gaussian units for all electromagnetic (EM) quantities]. This is also the structure of a Veselago Lens,
which will be discussed in Section \ref{section:calculation}  below.

This microstructure  is uniform in all $x,y$-planes, therefore all the eigenstates can have the form
$$ \langle{\bf r}|{\bf E}_n\rangle=
e^{i{\bf k}\cdot{\boldsymbol{\rho}
 }}{\bf f}(z),
$$
where {\bf k} is a real 2D wave vector in the $x,y$-plane while $\boldsymbol{\rho}$ is a 2D position
vector in that plane. It is easy to show that ${\bf f}(z)$ satisfies the following ordinary second
order differential equation
$$
\left(\frac{d^2}{dz^2}+k^2_2\right){\bf f}=u\theta_1({\bf r}){\bf f},
$$
where $\theta_1({\bf r})=1$ for $|z|<d$ and vanishes elsewhere. In each region of $z$
this is a one-dimensional Helmholtz equation,
the solution of which is a linear combination of sine and cosine functions with coefficients
that must satisfy $\nabla\cdot{\bf E}=0$ in each region. They must also satisfy
the outgoing wave condition for $|z|>d$ and continuity requirements
on $E_x$ and $E_y$ and $D_z\equiv\epsilon E_z$, as well as on all components of the
magnetic field, at the $\epsilon_1$, $\epsilon_2$ interfaces.
The microstructure is also invariant under the reflection $z\rightarrow -z$. These symmetries
are not violated by complex conjugation of the eigenfunctions. Therefore the eigenfunctions
can be characterized as transverse electric (TE) or transverse magnetic (TM), and also as
even $(+)$ or odd $(-)$ under $z\rightarrow -z$.

In the following subsections we will first find closed form expressions for the eigenstates
and closed form expressions for the nonlinear equation whose solutions are the eigenvalues.
In order to discuss qualitatively the properties of the eigenvalues we will restrict our
considerations to the case where $\epsilon_2$ is real and positive. In that case, when
 $\left|\mathbf{k}\right|<k_2,$ \mbox{$k_{2z}=\sqrt{k_{2}^{2}-\left|\mathbf{k}\right|^{2}}$} is real and the mode propagates out of the slab. However, when $\left|\mathbf{k}\right|>k_2,$ \mbox{$k_{2z}=\sqrt{k_{2}^{2}-\left|\mathbf{k}\right|^{2}}$} is imaginary and the mode decays away from the slab. We refer to these modes, respectively, as  propagating and evanescent modes. When $\left|\mathbf{k}\right|=k_2,$ $k_{2z}=0$ and the mode propagates parallel to the slab
without any decay or radiation away from the slab.

In Section \ref{subsection:Calculation_of_the_eigenvalues} we will describe
numerical calculations of the eigenvalues
for  the general non-quasistatic cases [Eqs.\ (\ref{eq:TM_res_eigen}), (\ref{eq:TE_res_eigen})]
and their consequences.
When the physical value of $\epsilon_1$ is very close to one of the eigenvalues
the contribution of this eigenstate to the physical
electric field can become very large, greatly exceeding the field in the absence of
the $\epsilon_1$ layer (Region II). When the physical $\epsilon_2\approx - \epsilon_1$ then
$s \approx 1/2$ and $s_k/\left(s-s_k\right)$ becomes very large for many of the TM eigenstates
when $k$ is large. This
may lead to a large contribution of the large $\left|\mathbf{k}\right|$ modes to the expansion of the electric field in Eq.\ (\ref{Eexpansion}), as already found earlier in the quasi-static regime \cite{BergPRA2014,FarhiBergVeselagoPRA2014}.

\subsection{The TM modes} 

Those are
\begin{eqnarray*}
\mathbf{E}_{\mathbf{k}}^{+}=e^{i\mathbf{k}\cdot\boldsymbol{\rho}}\left\{ \begin{array}{cc}
e^{-ik_{2z}z}A_{k}^{+}\left(\mathbf{e}_{z}\frac{k}{k_{2z}}+\mathbf{e}_{\mathbf{k}}\right) & \mathbf{r}\in\mathbf{\mathrm{I}}\\
B_{k}^{+}\left(-\mathbf{e}_{z}\frac{ik}{k_{1z}^{+}}\sin\left(k_{1z}^{+}z\right)+\mathbf{e}_{\mathbf{k}}\cos\left(k_{1z}^{+}z\right)\right) & \mathbf{r}\in\mathbf{\mathrm{II}}\\
e^{ik_{2z}z}A_{k}^{+}\left(-\mathbf{e}_{z}\frac{k}{k_{2z}}+\mathbf{e}_{\mathbf{k}}\right) & \mathbf{r}\in\mathbf{\mathrm{III}}
\end{array}\right.,\\\mathbf{E}_{\mathbf{k}}^{-}=e^{i\mathbf{k}\cdot\boldsymbol{\rho}}\left\{ \begin{array}{cc}
e^{-ik_{2z}z}A_{k}^{-}\left(\mathbf{e}_{z}\frac{k}{k_{2z}}+\mathbf{e}_{\mathbf{k}}\right) & \mathbf{r}\in\mathbf{\mathrm{I}}\\
B_{k}^{-}\left(\mathbf{e}_{z}\frac{ik}{k_{1z}^{-}}\cos\left(k_{1z}^{-}z\right)+\mathbf{e}_{\mathbf{k}}\sin\left(k_{1z}^{-}z\right)\right) & \mathbf{r}\in\mathbf{\mathrm{II}}\\
e^{ik_{2z}z}A_{k}^{-}\left(\mathbf{e}_{z}\frac{k}{k_{2z}}-\mathbf{e}_{\mathbf{k}}\right) & \mathbf{r}\in\mathbf{\mathrm{III}}
\end{array}\right..
\end{eqnarray*}
where \mbox{$\mathbf{e}_{\mathbf{k}}\equiv{\bf k}/|{\bf k}|,$} 
\mbox{$k_{1z}^\pm\equiv \sqrt{(k_1^\pm)^2-{\bf k}^2}$},\mbox{$k_{2z}\equiv \sqrt{(k_2)^2-{\bf k}^2}$}
\mbox{$k_1^\pm\equiv\sqrt{\epsilon^\pm_1}\omega/c,$} \mbox{$k_2\equiv\sqrt{\epsilon_2}\omega/c,$} and where $\nabla\cdot{\bf E}=0$
is already satisfied in Regions I, II, and III. Note that $E^+_z$ (i.e., the $z\mathrm{-component}$ of  $\mathbf{E}_{\mathbf{k}}^{+}$)
changes sign under the reflection $z\rightarrow -z$.
Thus, $E_z^+(-z)=-E_z^+(z)$ but $E_{\bf k}^+(-z)=E_{\bf k}^+(z)$ (this is the $x,y\mathrm{-plane}$ component of $\mathbf{E}_{\mathbf{k}}^{+}$)
in the even modes
while $E_z^-(-z)=E_z^-(z)$ but $E_{\bf k}^-(-z)=-E_{\bf k}^-(z)$ in the odd modes. For $k=0$ $\mathbf{e}_{\mathbf{k}}$ is not defined and we can replace it by $\mathbf{e}_{x}.$
The $A$ and $B$ coefficients are determined by the continuity requirements on
the tangential components of {\bf E} and the normal component of
${\bf D}\equiv\epsilon{\bf E}$ at the two interfaces $z=\pm d$. We thus get
\begin{eqnarray*}
B_{k}^{\pm}\left\{ \begin{array}{c}
\cos\left(k_{1z}^{+}d\right)\\
-\sin\left(k_{1z}^{-}d\right)
\end{array}\right\} =A_{k}^{\pm}e^{ik_{2z}d},\\
\frac{i\epsilon_{1}^{\pm}B_{k}^{\pm}}{k_{1z}^{\pm}}\left\{ \begin{array}{c}
\sin\left(k_{1z}^{+}d\right)\\
\cos\left(k_{1z}^{-}d\right)
\end{array}\right\} =\epsilon_{2}A_{k}^{\pm}\frac{e^{ik_{2z}d}}{k_{2z}}.
\end{eqnarray*}
From these two homogeneous linear equations for $A_k^\pm$ and $B_k^\pm$
we get the following nonlinear equation for the eigenvalues of $\epsilon_{1k}^\pm$
which also depend upon $k\equiv|{\bf k}|$ but not on the direction of {\bf k}:
\begin{equation}
\frac{\epsilon_{2}}{\epsilon_{1k}^{\pm}}=\frac{ik_{2z}}{k_{1z}^{\pm}}\left\{ \begin{array}{c}
\tan\left(dk_{1z}^{+}\right)\\
-\cot\left(dk_{1z}^{-}\right)
\end{array}\right\},
\label{eq:TM_res_eigen}
\end{equation}
From this equation it follows that the eigenvalues of $k_{1z}^{\pm}$ and
$\epsilon_{1k}^\pm$ depend only on the magnitude $k$ of the 2D wave vector {\bf k}.
The eigenfunctions depend on the direction of that vector only through the
$e^{i{\bf k}\cdot\boldsymbol{\rho}}$ factor.
When $\epsilon_2$ is real and positive and $k>k_{2}$ then  
$k_{2z}$ is imaginary and the eigenstate is evanescent and non-radiating.
Also, we can write
$k_{2z}\equiv i\kappa_{2z}$, leading to
the following form for the eigenvalue equation
$$
-\frac{\epsilon_{2}}{\epsilon_{1k}^{\pm}}=\frac{\kappa_{2z}}{k_{1z}^{\pm}}\left\{ \begin{array}{c}
\tan\left(dk_{1z}^{+}\right)\\
-\cot\left(dk_{1z}^{-}\right)
\end{array}\right\},
$$
From the above remark it follows that the solutions for $\epsilon_{1k}^{\pm}$
and $s_k^\pm$ are real
and involve no dissipation and no creation of energy. Many of those
$\epsilon_{1k}^{\pm}$ eigenvalues are negative and 
therefore the appropriate values of $s_k^\pm$ lie between 0 and 1,
as we found in the past for all the $s_k^\pm$ eigenvalues in the quasistatic
limit \cite{BergPRA2014}.

When $k$ is less than $k_{2}$ then $k_{2z}$ is real and the eigenstates will
be radiating energy away from the $\epsilon_1$ slab. In that case the eigenvalues
$k_{1z}^\pm$ will be complex and usually have real and imaginary parts, as can
be seen from Eq.\ (\ref{eq:TM_res_eigen}). In this case
$\epsilon_1^\pm$ must have an imaginary part with the ``wrong sign'' so as to
create energy that compensates for the radiation losses.

When $k$ is much larger than both $k_1^\pm$ and $k_{2}$ we get
$\kappa_{1z}^\pm\approx -k$, $\kappa_{2z}\approx -k$, and consequently
the eigenvalue equation becomes
$$
\frac{\epsilon_{2}}{\epsilon_{1k}^{\pm}}=-\left\{ \begin{array}{c}
\tanh\left(dk\right)\\
\coth\left(dk\right)
\end{array}\right\}.
$$
This agrees with results previously found in the quasi-static limit for all values of {\bf k}
 \cite{BergPRA2014,FarhiBergVeselagoPRA2014}. Clearly, when $dk\rightarrow\infty$
we get $\epsilon_{2}/\epsilon_{1k}^{\pm}\rightarrow -1$ or
$s_k^\pm\rightarrow 1/2$, which is therefore an accumulation point of the TM
eigenvalues.

The eigenfunction $\tilde{\bf E}_{\bf k}^\mp({\bf r})$, which is dual to
${\bf E}_{\bf k}^\mp({\bf r})$, is now chosen as
$$
\langle{\bf r}|{\bf E}_{{\bf k}}^{\mp}\rangle={\bf E}_{{\bf k}}^{\mp}({\bf r}),\;\langle{\bf r}|\tilde{{\bf E}}_{{\bf k}}^{\mp}\rangle=[{\bf E}_{-{\bf k}}^{\mp}({\bf r})]^{*},\,\langle\tilde{{\bf E}}_{{\bf k}}^{\mp}|{\bf r}\rangle=\langle{\bf r}|{\bf E}_{{-\bf k}}^{\mp}\rangle.
$$
It follows that any eigenstate is orthogonal to any dual eigenstate with a different
value of the 2D wave vector {\bf k}.

Even though $|{\bf E}_{\bf k}^-\rangle$ and $\langle\tilde{\bf E}_{\bf k}^+|$ are assured to be orthogonal because they usually have different eigenvalues, we also verified this by a direct
 calculation:
$$
\langle \tilde{\bf E}^+_{\bf k}|{\bf E}^-_{\bf k}\rangle=
 \int_{|z|<d} dV {\bf E}_{-{\bf k}}^+({\bf r})\cdot{\bf E}_{\bf k}^-({\bf r})=0.
$$
A similar direct calculation leads to
$\langle \tilde{\bf E}^-_{\bf k}|{\bf E}^+_{\bf k}\rangle=0.$

The  inner product of a TM mode and its dual leads to the following normalization integral
 ($L_x$, $L_y$ are the system sizes in the $x,y$-plane):
 \begin{widetext}
\begin{equation}
\label{eq:inner_product_TM}
\frac{\langle\tilde{{\bf E}}_{{\bf k}}^{\pm}|{\bf E}_{{\bf k}}^{\pm}\rangle}{L_{x}L_{y}}=\frac{\left(B_{k}^{\pm}\right)^{2}}{k_{1z}^{\pm2}}\left[-\left(k_{1}^{\pm}\right)^{2}d\pm\left(\left(k^{2}-k_{1z}^{\pm2}\right)\sin\left(2k_{1z}^{\pm}d\right)/2k_{1z}^{\pm}\right)\right].
%
%
\end{equation}
 \end{widetext}

\subsection{The TE modes}

Those are
$$\mathbf{E}_{\mathbf{k}}^{+}=e^{i\mathbf{k}\cdot\boldsymbol{\rho}}\left\{ \begin{array}{cc}
\mathbf{e}_{\perp}{A}^{+}_{\perp}e^{-ik_{2z}z} & \mathbf{r}\in\mathbf{\mathrm{I}}\\
\mathbf{e}_{\perp}{B}^{+}_{\perp}\cos\left(k^+_{1z}z\right) & \mathbf{r}\in\mathbf{\mathrm{II}}\\
\mathbf{e}_{\perp}{A}^{+}_{\perp}e^{ik_{2z}z} & \mathbf{r}\in\mathbf{\mathrm{III}}
\end{array}\right.
 $$
$$\mathbf{E}_{\mathbf{k}}^{-}=e^{i\mathbf{k}\cdot\boldsymbol{\rho}}\left\{ \begin{array}{cc}
-\mathbf{e}_{\perp}{A}^-_{\perp}e^{-ik_{2z}z} & \mathbf{r}\in\mathbf{\mathrm{I}}\\
\mathbf{e}_{\perp}{B}^-_{\perp}\sin\left(k^-_{1z}z\right) & \mathbf{r}\in\mathbf{\mathrm{II}}\\
\mathbf{e}_{\perp}{A}^-_{\perp}e^{ik_{2z}z} & \mathbf{r}\in\mathbf{\mathrm{III}}
\end{array}\right.
 $$

where ${\bf e}_\perp\equiv{\bf e}_{\bf k}\times{\bf e}_z$.
Note that ${E}_\perp$ is parallel to the slab and therefore does not change sign under the reflection
$z\rightarrow -z$.
Here $\nabla\cdot{\bf E}=0$ and $\nabla\cdot{\bf B}=0$ are satisfied automatically
in the various regions. However we need to impose the continuity of
${\bf E}\parallel{\bf e}_\perp$ and {\bf B}. This leads to
\begin{eqnarray*}
 {A}_{\perp}^{\pm}e^{ik_{2z}d}={B}_{\perp}^{\pm}\left\{ \begin{array}{c}
\cos\left(k_{1z}^{+}d\right)\\
\sin\left(k_{1z}^{-}d\right)
\end{array}\right\},\\
\mp ik_{2z}A_{\perp}^{\pm}e^{ik_{2z}d}=k_{1z}^{\pm}B_{\perp}^{\pm}\left\{ \begin{array}{c}
\sin\left(k_{1z}^{+}d\right)\\
\cos\left(k_{1z}^{-}d\right)
\end{array}\right\} .
\end{eqnarray*}
and hence to the following equation for the TE eigenvalues:
\begin{equation}
\frac{ik_{2z}}{k_{1z}^{\pm}}=\left\{ \begin{array}{c}
-\tan\left(k_{1z}^{+}d\right)\\
\cot\left(k_{1z}^{-}d\right)
\end{array}\right\}. 
\label{eq:TE_res_eigen}
\end{equation}

If $k<k_2$ then $k_{2z}$ is real and the eigenstates will be radiating states.
Solutions of Eq.\ (\ref{eq:TE_res_eigen}) for $k_{1z}^\pm$ will therefore have
a real part and an imaginary part and the eigenvalues $\epsilon_{1k}^\pm$
will have an imaginary part with the ``wrong sign''.

If $k>k_2$ then $k_{2z}=i\kappa_{2z}$ is imaginary and the eigenstates will
be evanescent non-radiating states. Therefore the $\epsilon_{1k}^{\pm}$
and $s_k^\pm$ eigenvalues are real
and involve no dissipation and no creation of energy.
Eq.\ (\ref{eq:TE_res_eigen}) then becomes
$$
-\frac{\kappa_{2z}}{k_{1z}^\pm}=\left\{ \begin{array}{c}
-\tan\left(dk_{1z}^{+}\right)\\
\cot\left(dk_{1z}^{-}\right)
\end{array}\right\}.
$$
Some consideration leads to the conclusion that all the solutions of this equation
for $k_{1z}^\pm$ will be either pure real, in which case
$\epsilon_{1k}^\pm>\epsilon_2$ and $s_k^\pm< 0$, or else
pure imaginary, in which case 
\mbox{$\epsilon_{1k}^+<\epsilon_2$} and \mbox{$0<s_k^+< 1$} while
\mbox{$\epsilon_{1k}^->\epsilon_2$} and \mbox{$s_k^-<0$}.
In the quasistatic limit, when \mbox{$k_2/k\rightarrow 0^+$}, we find that
\mbox{$\epsilon_2/\epsilon_1^\pm\rightarrow 0^-$} and therefore also
\mbox{$s_k^\pm\rightarrow 0^-$}. Consequently these states do not contribute
to the expansion of Eq.\ (\ref{Eexpansion})
for the local physical field.

The normalization integral of the TE modes is
\begin{eqnarray*}
\frac{ \langle\tilde{{\bf E}}_{{\bf k}}^{\pm}|{\bf E}_{{\bf k}}^{\pm}\rangle}{ L_xL_y}=\left(B_{\perp}^{\pm}\right)^{2}\left\{ \begin{array}{c}
\intop_{-d}^{d}dz\cos^{2}\left(k_{1z}^{+}z\right) \\
\intop_{-d}^{d}dz\sin^{2}\left(k_{1z}^{-}z\right)
\end{array}\right\}\\
=\frac{\left(B_{\perp}^{\pm}\right)^{2}}{2k_{1z}^{\pm}}\left[2k_{1z}^{\pm}d\pm\sin\left(2k_{1z}^{\pm}d\right)\right]\neq0.
\end{eqnarray*}


\subsection{Calculation of the eigenvalues}
\label{subsection:Calculation_of_the_eigenvalues}

The permittivity values and the slab thickness in the following calculations correspond to the values in the experiment with the PMMA-silver-photoresist setup described
in Ref. \cite{Zhang}, where in our case $\epsilon_1$ is the silver permittivity and $\epsilon_2$ is the average permittivity of PMMA and the photoresist. These permittivity values are appropriate for a free-space wavelength of $365\mathrm{nm}$.
We calculated the eigenvalues of the even and odd TM and TE modes according to Eqs. (\ref{eq:TM_res_eigen}) and  (\ref{eq:TE_res_eigen}), respectively, for $2d=35\mathrm{nm}, \lambda=365\mathrm{nm}$ and $\epsilon_2=2.57+0.09i$.
For any choice of $k\equiv\left|\mathbf{k}\right|$ there is an infinite number of solutions to the eigenvalue equations. Fortunately, for modes with high eigenvalues $\epsilon_{1k}$,  $s_k/\left(s-s_k\right) \rightarrow 0$ and these modes give a negligible contribution to the expansion of the electric field (see Eq. (\ref{Eexpansion})). We define these modes as the high order modes and associate low mode index numbers to the modes with low $\epsilon_{1\mathbf{k}}$ values.
\subsubsection{Eigenvalues of the TM modes}
\begin{figure}[h]
   \begin{center}
   \begin{tabular}{c}
   \includegraphics[width=8cm]{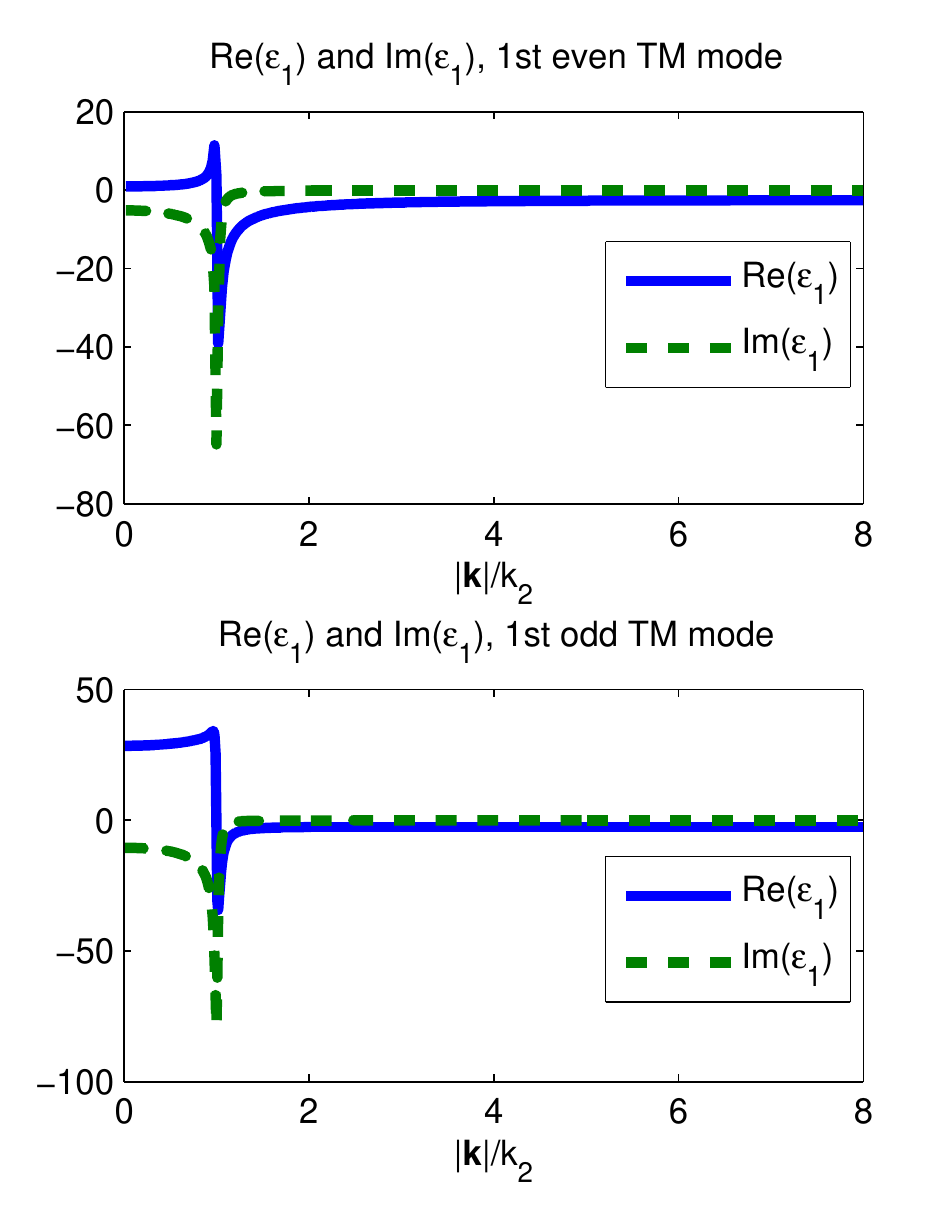}
   \end{tabular}
   \end{center}
   \caption[]
   { \label{fig:1st_modes_TM} 
The eigenvalues for the first even and odd TM modes as functions of $\left|\mathbf{k}\right|.$}
   \end{figure}

 In Fig. \ref{fig:1st_modes_TM} we present the eigenvalues of the first even and odd TM modes as functions of $\left|\mathbf{k}\right|$. It can be seen that in the limit $\left|\mathbf{k}\right| \rightarrow \infty$ the eigenvalues $\epsilon_{1\mathbf{k}}$ tend to $-\epsilon_2$ and hence $s_\mathbf{k} \thickapprox 1/2$. Thus, when the physical $\epsilon_2\thickapprox-\epsilon_1,$ $s\thickapprox 1/2$ and \mbox{$s_k/\left(s-s_k\right) \rightarrow \infty$}. Thus, the evanescent eigenstates which have spatial frequencies $\left|\mathbf{k}\right|>k_2$ play an important role in the imaging and can lead to an enhanced resolution image as argued in Ref. \cite{Pendry}. 

\begin{figure}[h]
   \begin{center}
   \begin{tabular}{c}
   \includegraphics[width=8cm]{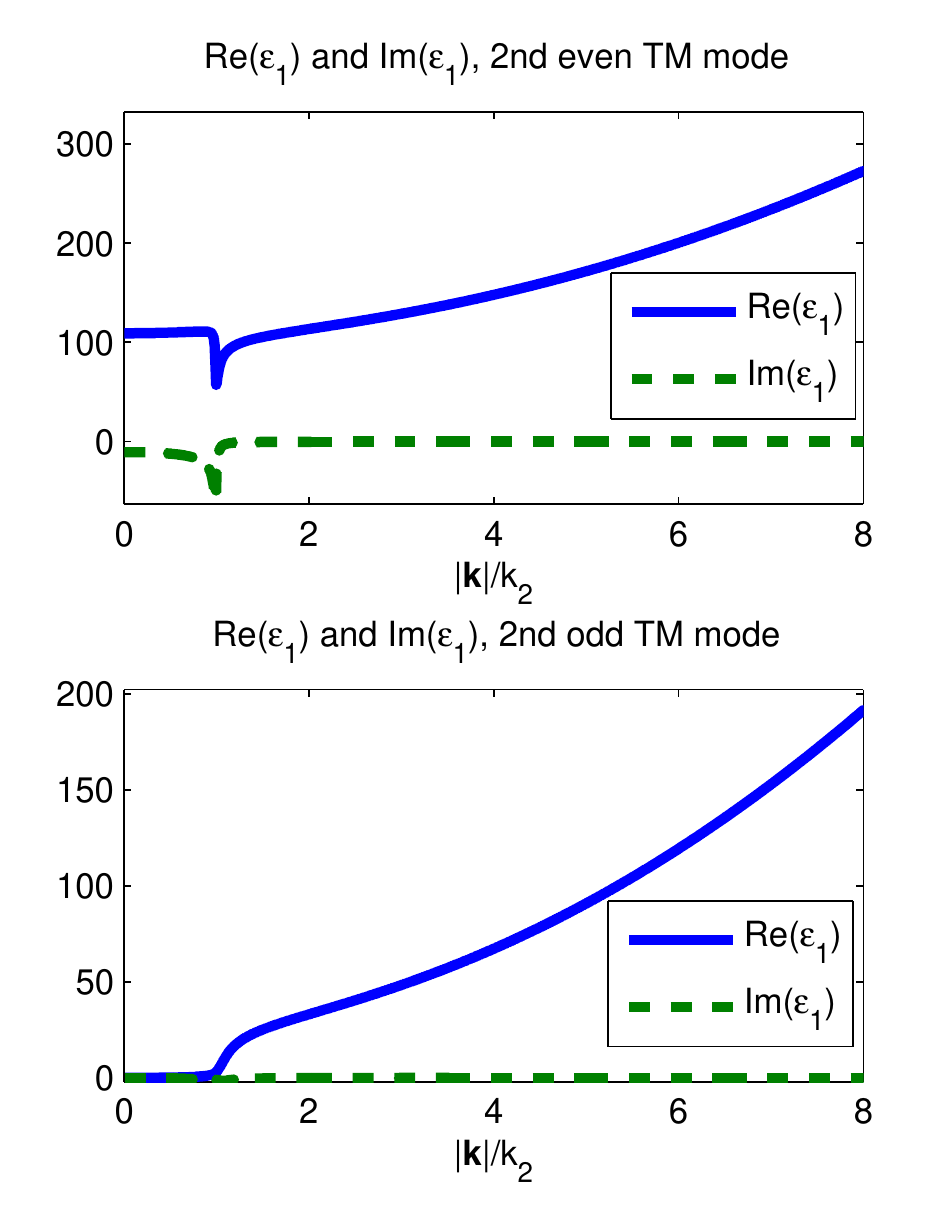}
   \end{tabular}
   \end{center}
   \caption[]
   { \label{fig:2nd_modes} 
The eigenvalues for the second even and odd TM modes as functions of $\left|\mathbf{k}\right|.$}
   \end{figure}

In Fig. \ref{fig:2nd_modes} we present the eigenvalues of the second even and odd TM modes as functions of $\left|\mathbf{k}\right|$. 
The second even mode has high values of $\mathrm{Re}\left(\epsilon_{1k}\right)$ which means that $s_k/\left(s-s_k\right) \approx 0$ and the contribution of this mode to the expansion is very small. Interestingly, the second odd mode, even though for large values of $\mathbf{\left|k\right|}$ has high values of $\mathrm{Re}\left(\epsilon_{1\mathbf{k}}\right)$, in the range where $\left|\mathbf{k}\right|\approx 0$ has $\epsilon_{1k}\approx 0$ which means that $s_k/\left(s-s_k\right)$ is not negligible for our physical $\epsilon_1$ and can become large for $\epsilon_1\approx 0$. 

Since the eigenstates do not decay in magnitude with time, one should expect that there should be constructive interference inside the slab. To verify this we calculated for the $\mathbf{k}=0$ eigenstates of the first two even and odd TM modes the phase accumulated due to the propagation in the $z$ and $-z$ directions inside the slab and the double reflection from the interfaces. The total phases for the round-trips inside the slab which were calculated were all integer multiples of $2\pi$ as expected.    

\subsubsection{Eigenvalues of the TE modes}
\begin{figure}[t]
   \begin{center}
   \begin{tabular}{c}
   \includegraphics[width=8cm]{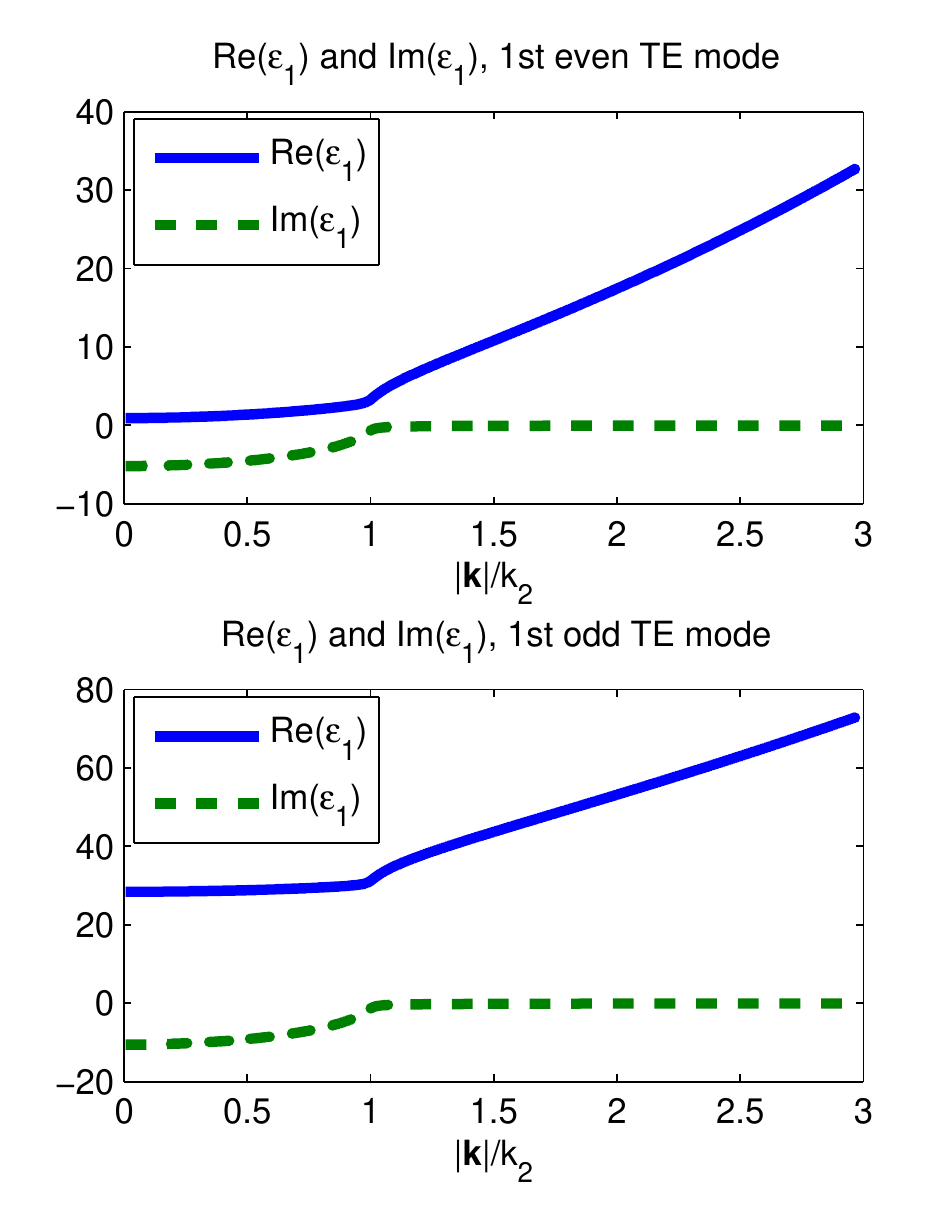}
   \end{tabular}
   \end{center}
   \caption[]
   { \label{fig:1st_modes_TE} 
The eigenvalues for the first even and odd TE modes as functions of $\left|\mathbf{k}\right|.$}
   \end{figure}

In Fig. \ref{fig:1st_modes_TE} we present the eigenvalues of the first even and odd TE modes as functions of $\left|\mathbf{k}\right|$. It can be seen that in the limit $\left|\mathbf{k}\right| \rightarrow \infty$ the eigenvalues $\epsilon_{1\mathbf{k}}$ tend to infinity. When $\left|\mathbf{k}\right|=0$ the modes propagate perpendicular to the slab and can therefore be defined both as TM and TE. This is apparent in the equality of the TM and TE eigenvalues at $\left|\mathbf{k}\right|=0$.

\begin{figure}[h]
   \begin{center}
   \begin{tabular}{c}
   \includegraphics[width=8cm]{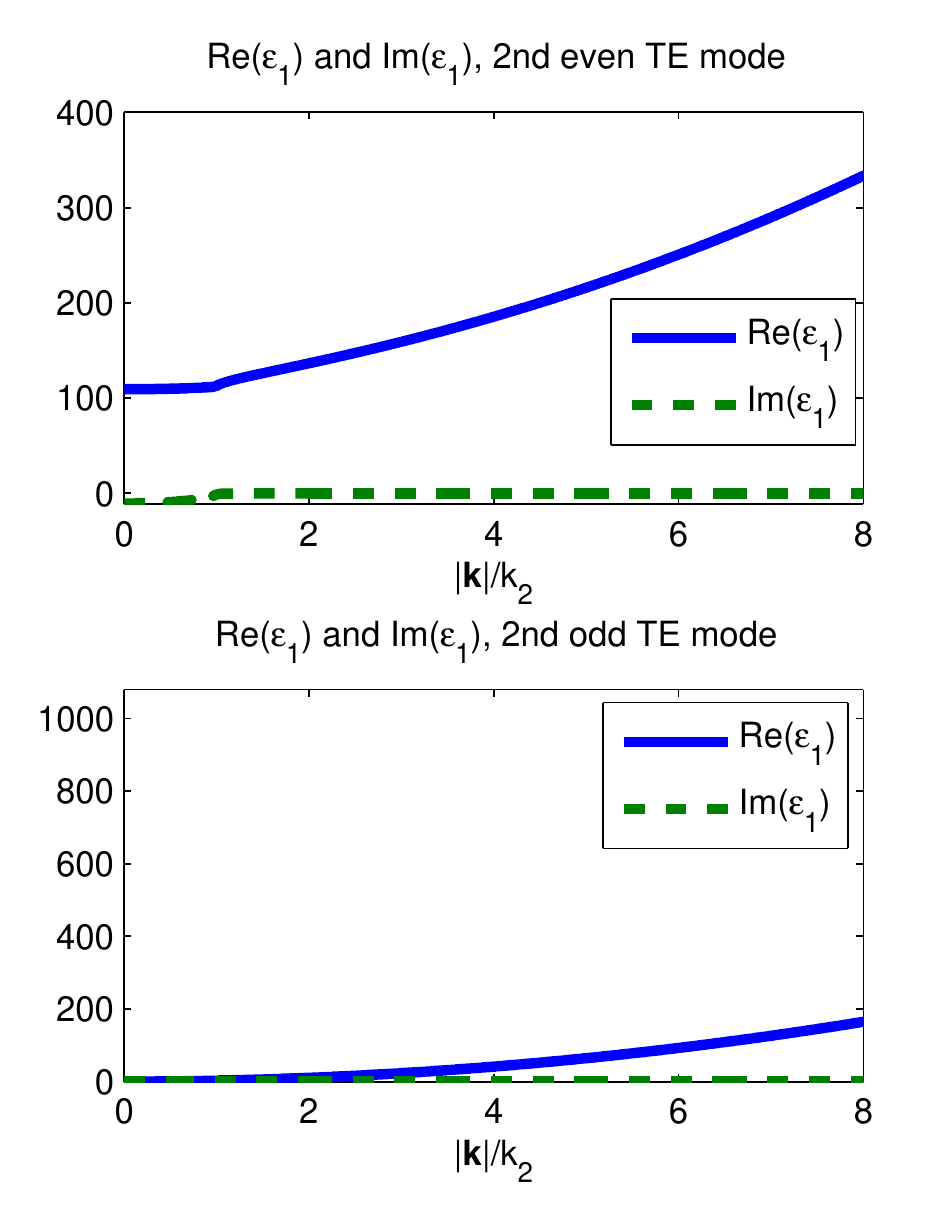}
   \end{tabular}
   \end{center}
   \caption[]
   { \label{fig:2nd_modes_TE} 
The eigenvalues for the second even and odd TE modes as functions of $\left|\mathbf{k}\right|.$}
   \end{figure}

In Fig. \ref{fig:2nd_modes_TE} we present the eigenvalues of the second even and odd TE modes as functions of $\left|\mathbf{k}\right|$. Here too, the eigenvalues at $\left|\mathbf{k}\right|=0$ are the same as those of the TM modes. It can also be seen that the first even and second odd TE modes at $\left|\mathbf{k}\right|\approx0$ can give a small and a significant contribution to the expansion of $\mathbf{E}\left(\mathbf{r}\right)$, respectively, since their $\epsilon_{1k}$ values are not far from physical values of $\epsilon_1.$ On the other hand, the first odd and second even TE modes should give a negligible contribution to the expansion since their $\epsilon_{1k}$ values are far from physical values of $\epsilon_1.$ Interestingly, the eigenvalues of the second odd TE mode have a small imaginary part and are close to physical $\epsilon_1$ values which are realizable in experiments.

\section{Using the eigenstate expansion to calculate the electric field of a point electric dipole
in a flat slabs structure}
\label{section:calculation}

We now  use the eigenfunctions derived in  Section \ref{FlatSlabs} to expand the resulting electric field. Fig.\ \ref{fig:capacitor} shows
this structure, where the object and images according to geometrical optics are represented by circles. We will consider oscillating electric point dipoles 
in Region III directed along $z$ and $x$ axes as the source of the EM field.

\subsection{Dipole object directed along $z$}


We consider an oscillating electric point dipole at \mbox{${\bf r}=(0,0,z_0)\equiv{\bf z}_0$}
in Region III directed
along $z$ as the source of the EM field. The current distribution of the dipole at $\mathbf{z}_0$ can be written as \mbox{$\mathbf{J}_{\mathrm{dip}}=-i\mathbf{e}_{z}
 \omega p\delta^3\left(\mathbf{r}-\mathbf{z}_{0}\right)$} where $p$ is the \emph{electric dipole moment}. 

The electric field of this dipole in a uniform $\epsilon_2$ medium is
\begin{widetext}
\begin{equation}
\label{eq:dipole_uniform_medium}
\mathbf{E}_{0}\left(\mathbf{r}\right)=\frac{1}{\epsilon_{2}}e^{ik_{2}r}\left\{ \left[k_{2}^{2}\left(\mathbf{n}\times\mathbf{p}\right)\times\mathbf{n}\right]\frac{1}{r}+\left[3\mathbf{n}\left(\mathbf{n}\cdot\mathbf{p}\right)-\mathbf{p}\right]\left(\frac{1}{r^{3}}-\frac{ik_{2}}{r^{2}}\right)\right\} ,
\end{equation}
\end{widetext}
where $r\equiv\left|\mathbf{r}-\mathbf{z}_{0}\right|,\mathbf{n}\equiv\frac{\mathbf{r}-\mathbf{z}_{0}}{\left|\mathbf{r}-\mathbf{z}_{0}\right|}.$
This  differs from the expression for the electric field of an electric point dipole in vacuum \cite{JacksonBook1975}  by the $\epsilon_2$ factor in the denominator and by the appearance
 of $k_{2}\equiv\sqrt{\epsilon_2}\omega/c$ instead of just $\omega/c$.

Because the expression for ${\bf E}_0({\bf r})$  can be obtained by using Green's
function of Eq.\ (\ref{Gspherical})
and ${\bf J}_{\rm{ dip}}$ defined above,
therefore the scalar product $\langle\tilde{\bf E}^\pm_{\bf k}|{\bf E}_0\rangle$, which appears
in Eq.\ (\ref{Eexpansion}), can be written as
\begin{eqnarray}
\langle{\bf \tilde{E}}^\pm_{\mathbf{k}}|{\bf E}_{0}\rangle
&=&  -\,\frac{4\pi i}{\epsilon_{2}\omega} \int dV'\theta_{1}\left(\mathbf{r}'\right)
{\bf{E}}^\pm_{-\mathbf{k}}\left(\mathbf{r}'\right)\cdot\nonumber\\ 
&&\hspace{5 mm}\cdot\int dV\overleftrightarrow{G}\left(\mathbf{r}',
\mathbf{r}\right)\cdot\mathbf{J}_{\rm{ dip}}\left(\mathbf{r}\right)\nonumber \\
&=&-\frac{4\pi i}{\epsilon_{2}\omega} s^\pm_k\int dV{\bf {E}}^\pm_{-\mathbf{k}}\left(\mathbf{r}\right)\cdot
\mathbf{J}_{\rm {dip}}\left(\mathbf{r}\right)\hspace{10 mm}
\label{Jintegral} \\
  &=&-\frac{4\pi p}{\epsilon_2}s^\pm_k E^\pm_{-{\bf k}z}({\bf z}_0).
\label{eq:simplify}
\end{eqnarray}
It can be seen that an oscillating electric dipole introduces in $\langle{\bf \tilde{E}}^\pm_{\mathbf{k}}|{\bf E}_{0}\rangle$ an additional factor $s_k$ which leads to a singularity when $\epsilon_{1\bf k}=\epsilon_2$. This differs from the $\epsilon_{1\bf k}=\epsilon_1$ singularity which arises from  $s_k/\left(s-s_k\right).$
 The inner product vanishes for all of the TE modes, but for each of the TM modes we get

\begin{eqnarray}
\frac{\left\langle \mathbf{\tilde{E}_{k\,\mathrm{TM}}^{\pm}}|\mathbf{E}_{0}\right\rangle }{L_{x}L_{y}}=B_{k}^{\pm}\left\{ \begin{array}{c}
\cos(dk_{1z}^{+})\\
\sin(dk_{1z}^{-})
\end{array}\right\} \frac{4\pi ps_{k}^{\pm}ke^{ik_{2z}(z_{0}-d)}}{\epsilon_{2}k_{2z}}.
\label{eq:inner_Ek_E0}
 \end{eqnarray}
Eq.\ (\ref{Eexpansion}) now becomes
\be
{\bf E}({\bf r})-{\bf E}_{0}({\bf r})=\sum_{{\rm TM},\,\alpha=\pm}\int\frac{d^{2}k}{\left(2\pi\right)^{2}}\frac{s_{k}^{\alpha}}{s-s_{k}^{\alpha}}\frac{\langle\tilde{{\bf E}}_{\mathbf{k}}^{\alpha}|{\bf E}_{0}\rangle}{\langle\tilde{{\bf E}}_{\mathbf{k}}^{\alpha}|{\bf E}_{\mathbf{k}}^{\alpha}\rangle}{\bf E}_{{\bf k}}^{\alpha}({\bf r}), \label{DipoleFieldExpansion}
\ee
where  $\sum_{{\rm TM},\,\alpha=\pm}$ means that one should sum over all the TM $(+)$
and TM $(-)$ eigenstates.
Since the only dependence on the direction of {\bf k} comes from the unit vector ${\bf e_k}$
and the factor $e^{i{\bf k}\cdot{\mathbf\rho}}$ which are in ${\bf E}_{\bf k}^\alpha({\bf r})$,
therefore the integration over the azimuth angle $\varphi$ between {\bf k} and $\boldsymbol{\rho}$ can be carried out analytically,
as we show in Section \ref{Analytic} below. This leaves only a 1D integration over $k\equiv|{\bf k}|$ to be calculated numerically.
Those integrals are shown below.

One can see from Eq. (\ref{Jintegral}) that when the source is located far from the slab, the evanescent eigenstates give only a small contribution to Eq. (\ref{DipoleFieldExpansion})
 since they have a small amplitude at that location. This is apparent from the exponential factor in Eq. (\ref{eq:inner_Ek_E0}) which expresses this evanescent behavior.

\subsubsection{Analytic integration with respect to $\varphi$}
\label{Analytic}

We notice that in the eigenstate expansion the only expression which depends on the 2D orientation of $\mathbf{k}$ is $|{\bf E}_{{\bf k}}^{\pm}\rangle$.
Therefore the integration over the azimuthal angle $\varphi$ can be performed analytically
\begin{widetext}
\begin{equation}
\int\mathbf{E}_{\mathbf{k}}^{\pm}d\varphi=2\pi B_{k}^{\pm}\left\{ \begin{array}{cc}
e^{-ik_{2z}\left(z+d\right)}\left\{ \begin{array}{c}
\cos\left(k_{1z}^{+}d\right)\\
\sin\left(k_{1z}^{-}d\right)
\end{array}\right\} \left(\pm\mathbf{e}_{z}J_{0}\left(k\rho\right)\frac{k}{k_{2z}}\pm i\mathbf{e_{\boldsymbol{\rho}}}J_{1}\left(k\rho\right)\right) & \mathbf{r}\in\mathbf{\mathrm{I}}\\
\mathbf{e}_{z}\frac{ik}{k_{1z}^{\pm}}\left\{ \begin{array}{c}
-\sin\left(k_{1z}^{+}z\right)\\
\cos\left(k_{1z}^{-}z\right)
\end{array}\right\} J_{0}\left(k\rho\right)+\mathbf{e_{\boldsymbol{\rho}}}iJ_{1}\left(k\rho\right)\left\{ \begin{array}{c}
\cos\left(k_{1z}^{+}z\right)\\
\sin\left(k_{1z}^{-}z\right)
\end{array}\right\}  & \mathbf{r}\in\mathbf{\mathrm{II}}\\
e^{ik_{2z}\left(z-d\right)}\left\{ \begin{array}{c}
\cos\left(k_{1z}^{+}d\right)\\
\sin\left(k_{1z}^{-}d\right)
\end{array}\right\} \left(-\mathbf{e}_{z}J_{0}\left(k\rho\right)\frac{k}{k_{2z}}+\mathbf{e_{\boldsymbol{\rho}}}iJ_{1}\left(k\rho\right)\right) & \mathbf{r}\in\mathbf{\mathrm{III}}
\end{array}\right.,
\label{eq:after_phi_integration}
\end{equation}
\end{widetext}
where $J_{0}\left(x\right)$ is a Bessel function of the first kind.
Eq. (\ref{DipoleFieldExpansion}) now reads
\begin{align}
\label{Eexpansion3}
&|{\bf E}\rangle-|{\bf E}_{0}\rangle\nonumber \\
&=\sum_{\mathrm{TM}}\sum_{+,-}\int\frac{dk}{\left(2\pi\right)^{2}}\frac{s_{\mathbf{k}}}{s-s_{\mathbf{k}}}\frac{\langle\tilde{{\bf E}}_{\mathbf{k}}|{\bf E}_{0}\rangle}{\langle\tilde{{\bf E}}_{\mathbf{k}}|{\bf E}_{\mathbf{k}}\rangle}\left(\int|{\bf E}_{\mathbf{k}}\rangle d\varphi\right)k \nonumber \\
&\equiv\sum_{\mathrm{TM}}\sum_{+,-}\int dk\mathbf{F}\left(\mathbf{r},k\right),
 \end{align}
where $\left(\int|{\bf E}_{\mathbf{k}}\rangle d\varphi\right)$ is given by Eq. (\ref{eq:after_phi_integration}).

\subsubsection{Calculation of the integrands as functions of $\left|\mathbf{k}\right|$}
\begin{figure}[t]
   \begin{center}
   \begin{tabular}{c}
   \includegraphics[width=9cm]{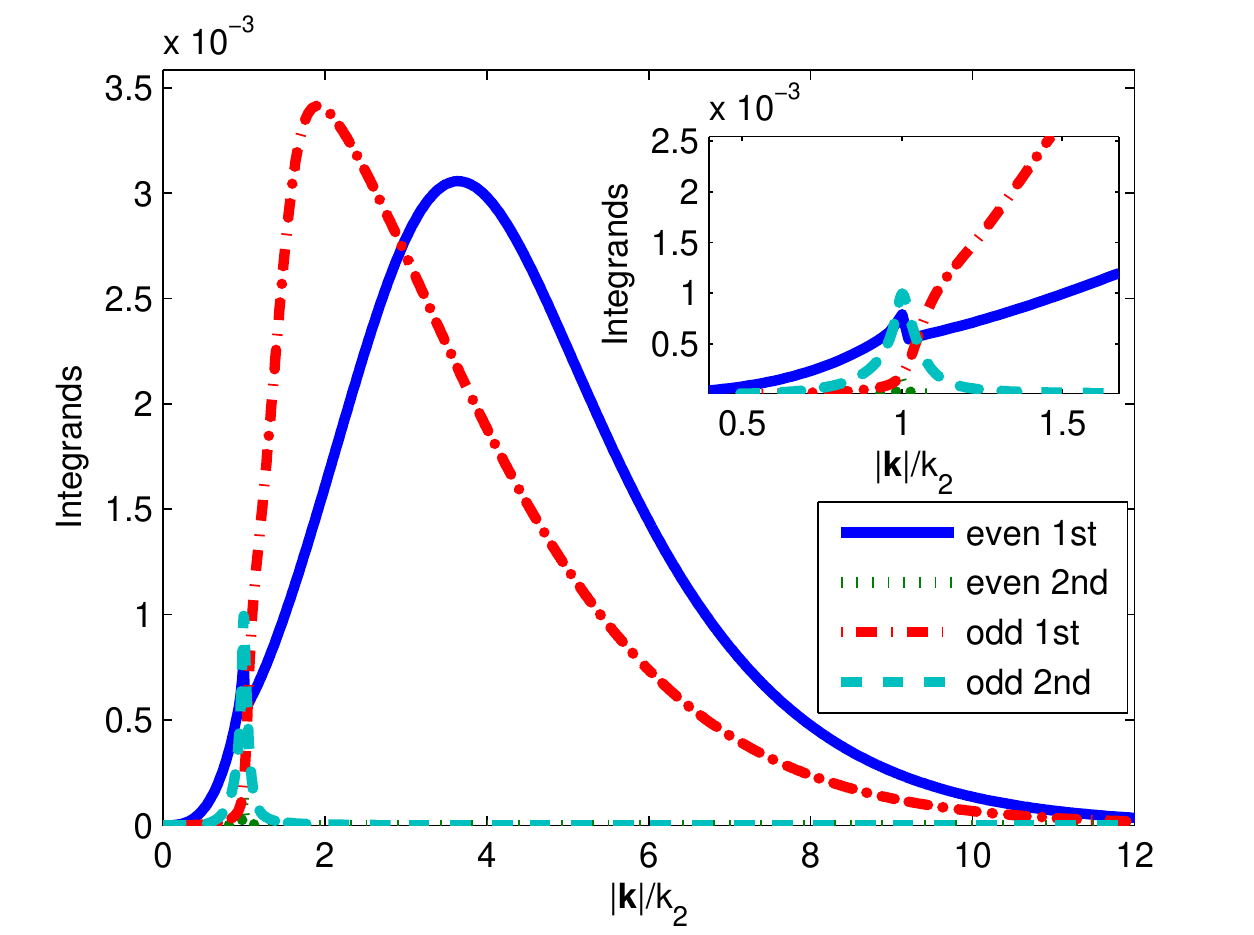}
   \end{tabular}
   \end{center}
   \caption[]
   { \label{fig:integrands} 
Integrands of Eq.  (\ref{Eexpansion3}) for the first two even and odd TM modes at $\rho=0,z=-d.$}
   \end{figure}

We calculated the integrands in Eq.\ (\ref{Eexpansion3}) for the first two even and odd TM modes for the coordinates $z=-d,\rho=0.$ This was performed by simply substituting the physical parameters and the eigenvalues in $s_\mathbf{k}/\left(s-s_\mathbf{k}\right)$ and in  Eqs. (\ref{eq:inner_product_TM}),(\ref{eq:inner_Ek_E0}) and (\ref{eq:after_phi_integration}), where in the last expression we also substituted the coordinates. In Fig. \ref{fig:integrands} we present the integrands as functions of $\left|\mathbf{k}\right|$. It can be seen that the modes with the dominant contribution to the expansion are the first even and odd modes. The second even and odd modes give a negligible contribution and a very small contribution to the expansion, respectively (very small values of the integrand of the second even mode which cannot be seen in the figure). This validates our analysis in the previous subsection.

\subsubsection{Calculation of the electric field}

\begin{figure}[t]
   \begin{center}
   \begin{tabular}{c}
   \includegraphics[height=6cm]{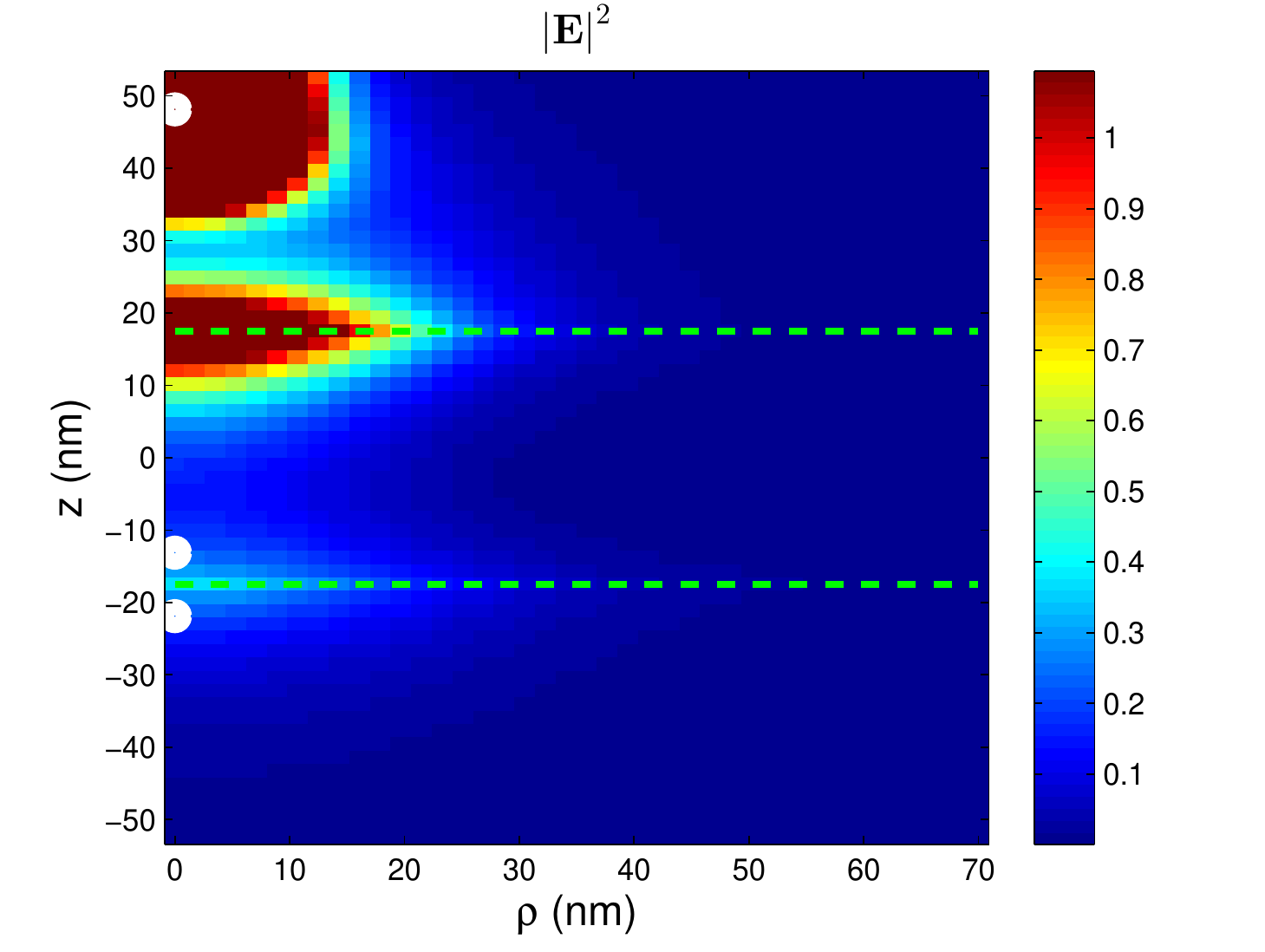}
   \end{tabular}
   \end{center}
   \caption[]
   { \label{fig:intensity} 
$\left|\mathbf{E}\right|^{2}$ in the three regions for a dipole located at $\rho=0,$ $z=d+7\cdot\left(2d\right)/8$  and $\epsilon_1=-2.55+0.24i, \epsilon_2=2.57+0.09i.$}
  \end{figure}

\paragraph{A setup with $\epsilon_{2}=2.57+0.09i,\epsilon_{1}=-2.55+0.24i$}
We calculated the electric field in the three regions by numerically integrating Eq. (\ref{Eexpansion3}) with respect to $\left|\mathbf{k}\right|.$  In Fig. \ref{fig:intensity} we present $\left|\mathbf{E}\right|^{2}$ in the three regions for a dipole located at $z=d+7\cdot\left(2d\right)
 /8$ and permittivity values of $\epsilon_1=-2.55+0.24i, \epsilon_2=2.57+0.09i.$  The white circles denote the object and the image expected according
to geometrical optics. In this
figure, as well as in Fig. \ref{fig:intensity2} that display all
the regions, we used a linear color scale. In order to
present an informative figure we mapped all the values higher than a certain value to this value. Thus, in all
the locations which exhibit the highest value, the actual
values are often much higher than the apparent value. It can be seen that the maximal intensity is at the interfaces between the slab and the surrounding medium.
\begin{figure}[t]
   \begin{center}
   \begin{tabular}{c}
   \includegraphics[width=7cm]{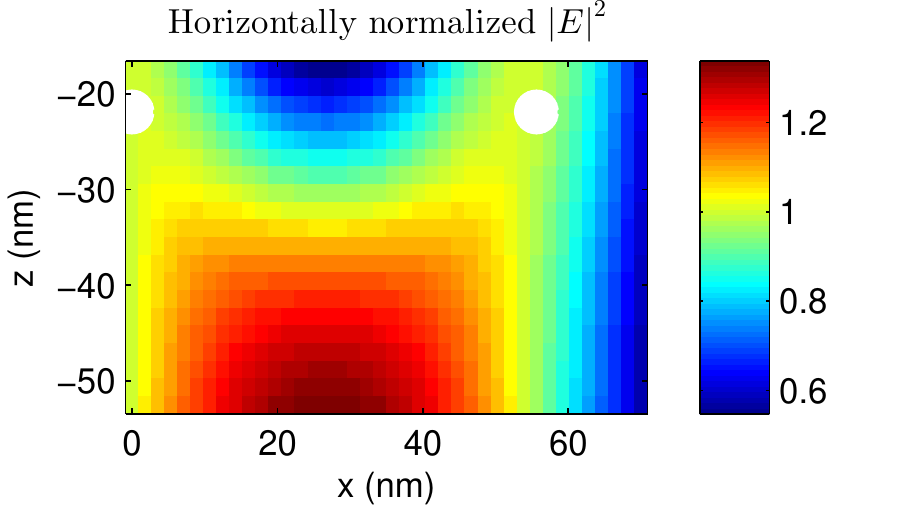}
   \end{tabular}
   \end{center}
   \caption[]
   { \label{fig:resolution} 
Horizontally normalized $\left|\mathbf{E}\right|^{2}$ in Region I for two oscillating dipoles located at $z=d+7\cdot\left(2d\right)/8$ and \mbox{$\epsilon_1=-2.55+0.24i,$} \mbox{$\epsilon_2=2.57+0.09i.$}}
  \end{figure} 
In Fig. \ref{fig:resolution} we present  $\left|\mathbf{E}\right|^{2}$, normalized by the maximal horizontal intensity, in Region I for two horizontally distanced electric dipole sources. The location of the second dipole was set to be such that the field intensity at the midpoint between the two images is $e^{-1/2}$ times the intensity at the image maximum. We define this distance
as the separation distance needed to resolve the two images. It can be seen that the optimal resolution is at the interface between the slab and the medium. Thus, the optimal imaging is at the interface between the slab and Region I in terms of both intensity and resolution. 
   These results are in agreement with our quasistatic analysis in Ref. \cite{FarhiBergVeselagoPRA2014}.

\paragraph{A setup in which $s-1/2$ is divided by 1000}
We divided $\Delta s\equiv s - \frac{1}{2}$ by $1000$ and calculated the electric field in the three regions. This setup, in which $\epsilon_1$ is much closer to $-\epsilon_2$, was expected to achieve better resolution according to the explanation in Subsection \ref{subsection:Calculation_of_the_eigenvalues}.
In Fig.  \ref{fig:intensity2} we present $\left|\mathbf{E}\right|^{2}$ in the three regions for a dipole located at $z=d+7\cdot\left(2d\right)/8$ (top region) and $\Delta s$ divided by $1000$. It can be seen that the intensity here is higher by more than an order of magnitude compared to the previous case.
\begin{figure}[t]
   \begin{center}
   \begin{tabular}{c}
   \includegraphics[height=6cm]{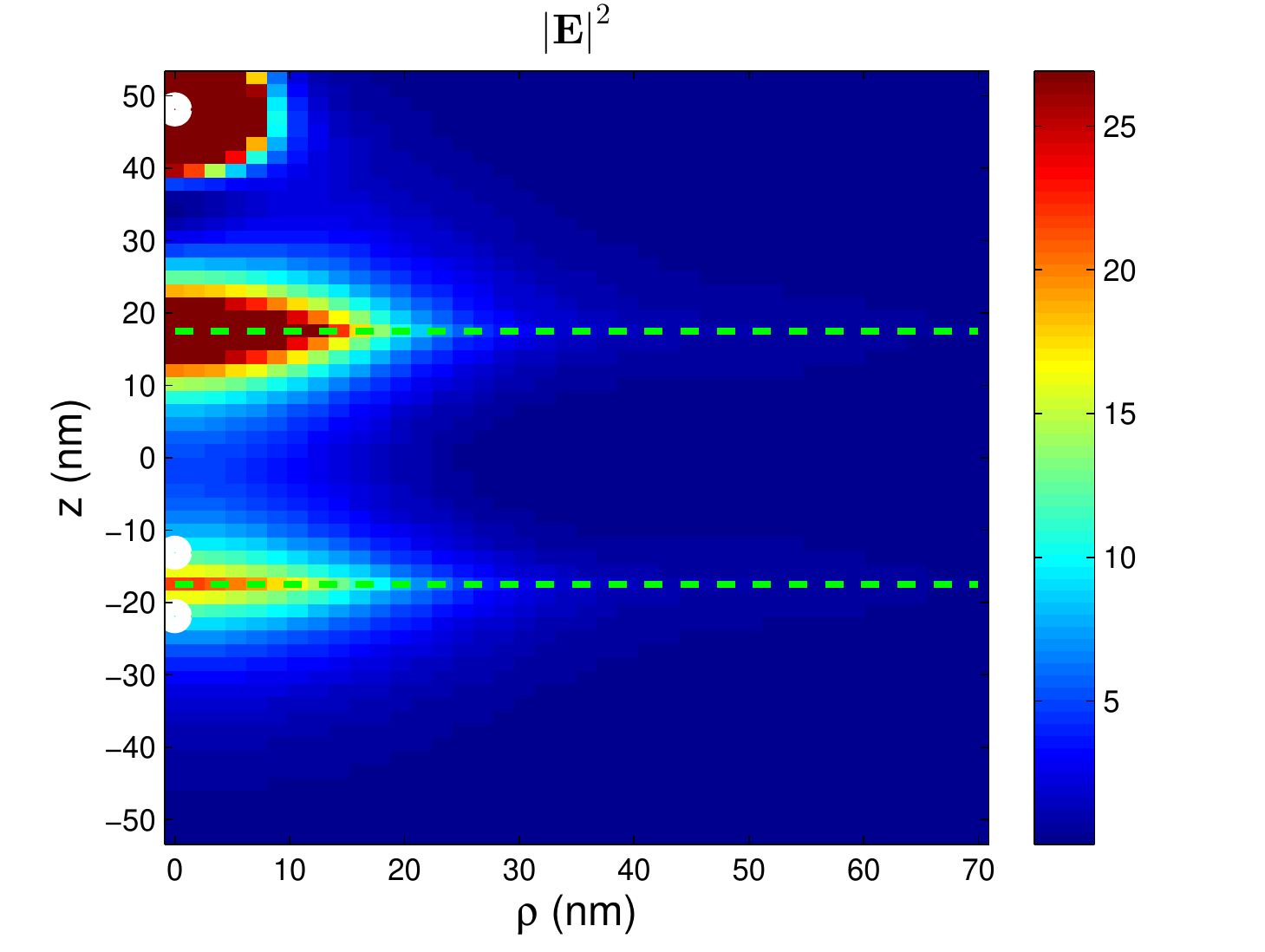}
   \end{tabular}
   \end{center}
   \caption[]
   { \label{fig:intensity2} 
$\left|\mathbf{E}\right|^{2}$ in the three regions for a dipole located at \mbox{$z=d+7\cdot\left(2d\right)/8$} and $\Delta s$ divided by $1000.$}
  \end{figure} 
\begin{figure}[t]
   \begin{center}
   \begin{tabular}{c}
   \includegraphics[width=7cm]{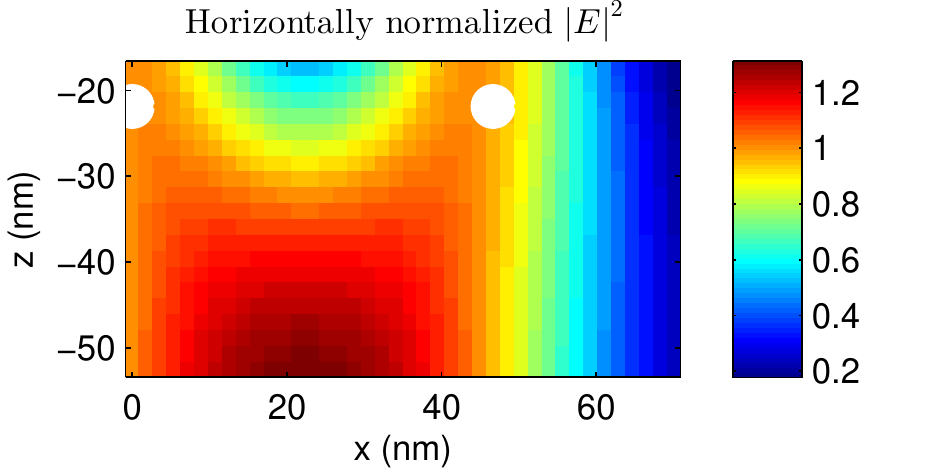}
   \end{tabular}
   \end{center}
   \caption[]
   { \label{fig:resolution2} 
Horizontally normalized $\left|\mathbf{E}\right|^{2}$ in Region I for two oscillating dipoles located at $z=d+7\cdot\left(2d\right)/8$ and $\Delta s$ divided by $1000$.}
  \end{figure} 
In Fig. \ref{fig:resolution2} we present the horizontally normalized $\left|\mathbf{E}\right|^{2}$ in Region I for two horizontally distanced dipoles. Here, too, the optimal resolution is at the interface between the slab and Region I. 

In conclusion, when we decrease $\Delta s,\,$ $\epsilon_1$ becomes closer to $\epsilon_{1\mathbf{k}}$ for the evanescent modes. Thus, there is a stronger amplification of these modes and the resolution is further enhanced since modes with higher $\mathbf{\left|k\right|}$ values are exploited in the imaging. It should be noted that $\Delta s \thickapprox 0$ can be achieved also when the imaginary parts of $\epsilon_1$ and $\epsilon_2$ have opposite signs, in which case one of the constituents exhibits dissipation while the other exhibits gain.

\subsection{Dipole object directed along $x$}

We consider an oscillating electric point dipole at \mbox{${\bf r}=(0,0,z_0)\equiv{\bf z}_0$}
in Region III directed 
along $x$ as the source of the EM field. The current distribution of the dipole at $\mathbf{z}_0$ can be written as \mbox{$\mathbf{J}_{\mathrm{dip}}=-i\mathbf{e}_{x}\omega p\delta^3\left(\mathbf{r}-\mathbf{z}_{0}\right)$} where $p$ is the \emph{electric dipole moment}. 
The electric field of this dipole in a uniform $\epsilon_2$ medium is given by the expression in Eq. (\ref{eq:dipole_uniform_medium}), where $\mathbf{p}=p\mathbf{e}_{x}.$

\subsubsection{Contribution of the TM modes}
We define $\varphi_{\mathbf{k}}$ as the azimuthal angle of $\mathbf{k}$ relative to $\mathbf{e}_{x}$ and project 
$\mathbf{e}_{\mathbf{k}}$ onto the $x$ and $y$ axes. We substitute \mbox{$\left(\mathbf{e}_{\mathbf{k}}\cdot\mathbf{e}_{x}\right)=\cos\varphi_{\mathbf{k}},\,\left(\mathbf{e}_{\mathbf{k}}\cdot\mathbf{e}_{y}\right)=\sin\varphi_{\mathbf{k}}$} and write the TM eigenfunctions as follows
\begin{widetext}
\begin{equation}
\mathbf{E}_{\mathbf{k}}^{\pm}=B_{k}^{\pm}e^{i\mathbf{k}\cdot\boldsymbol{\rho}}\left\{ \begin{array}{cc}
\pm e^{-ik_{2z}\left(z+d\right)}\cos\left(k_{1z}^{\pm}d\right)\left(\mathbf{e}_{z}\frac{k}{k_{2z}}+\mathbf{e}_{x}\cos\varphi_{\mathbf{k}}+\mathbf{e}_{y}\sin\varphi_{\mathbf{k}}\right) & \mathbf{r}\in\mathbf{\mathrm{I}}\\
\mp\mathbf{e}_{z}\frac{ik}{k_{1z}^{+}}\sin\left(k_{1z}^{\pm}z\right)+\left(\mathbf{e}_{x}\cos\varphi_{\mathbf{k}}+\mathbf{e}_{y}\sin\varphi_{\mathbf{k}}\right)\cos\left(k_{1z}^{+}z\right) & \mathbf{r}\in\mathbf{\mathrm{II}}\\
e^{ik_{2z}\left(z-d\right)}\cos\left(k_{1z}^{\pm}d\right)\left(-\mathbf{e}_{z}\frac{k}{k_{2z}}+\mathbf{e}_{x}\cos\varphi_{\mathbf{k}}+\mathbf{e}_{y}\sin\varphi_{\mathbf{k}}\right) & \mathbf{r}\in\mathbf{\mathrm{III}}
\end{array}\right..
\end{equation}
 \end{widetext}
 The scalar product $\left\langle \mathbf{\tilde{E}_{k\,\mathrm{TM}}^{\pm}}|\mathbf{E}_{0}\right\rangle$ can be written according to Eq. (\ref{Jintegral}) as
\begin{align}
&\frac{\left\langle \mathbf{\tilde{E}_{k\,\mathrm{TM}}^{+}}|\mathbf{E}_{0}\right\rangle }{L_{x}L_{y}}\nonumber\\
&=\frac{4\pi ps_{k}^{\pm}}{\epsilon_{2}}B_{k}^{\pm}e^{ik_{2z}\left(z_{0}-d\right)}\left\{ \begin{array}{c}
\cos\left(k_{1z}^{+}d\right)\\
\sin\left(k_{1z}^{-}d\right)
\end{array}\right\} \cos\left(\varphi_{\mathbf{k}}\right).
\end{align}

Here both $\left\langle \mathbf{\tilde{E}_{k\,\mathrm{TM}}^{\pm}}|\mathbf{E}_{0}\right\rangle$ and $\mathbf{E}_{\mathbf{k}}^{\pm}$ depend upon $\varphi_\mathbf{k}.$ We change the integration variables $d^{2}k=kd\varphi_{\mathbf{k}}dk$
and integrate $\frac{\left\langle \mathbf{\tilde{E}_{k\,\mathrm{TM}}^{\pm}}|\mathbf{E}_{0}\right\rangle }{L_{x}L_{y}}\mathbf{E}_{\mathbf{k}}^{\pm}$ analytically with respect to the azimuthal angle $\varphi_\mathbf{k}$
\begin{widetext}
\begin{gather}
\int\frac{\left\langle \mathbf{\tilde{E}_{k\,\mathrm{TM}}^{\pm}}|\mathbf{E}_{0}\right\rangle }{L_{x}L_{y}}\mathbf{E}_{\mathbf{k}}^{\pm}d\varphi_\mathbf{k}=\frac{4\pi ps_{k}^{\pm}}{\epsilon_{2}}e^{ik_{2z}\left(z_{0}-d\right)}\left\{ \begin{array}{c}
\cos\left(k_{1z}^{+}d\right)\\
\sin\left(k_{1z}^{-}d\right)
\end{array}\right\} 2\pi\left(B_{k}^{\pm}\right)^{2}\times\nonumber \\
\left\{ \begin{array}{cc}
e^{-ik_{2z}\left(z+d\right)}\left\{ \begin{array}{c}
\cos\left(k_{1z}^{+}d\right)\\
-\sin\left(k_{1z}^{-}d\right)
\end{array}\right\} \left(\mathbf{e}_{z}\frac{k}{k_{2z}}i\cos\left(\varphi_{\boldsymbol{\rho}}\right)J_{1}(k\rho)+\mathbf{e}_{x}\left[\cos^{2}\left(\varphi_{\boldsymbol{\rho}}\right)J_{0}(k\text{\ensuremath{\rho}})-\frac{\cos\left(2\varphi_{\boldsymbol{\rho}}\right)J_{1}(k\rho)}{k\text{\ensuremath{\rho}}}\right]-\mathbf{e}_{y}\frac{\sin\left(2\varphi_{\boldsymbol{\rho}}\right)}{2}J_{2}(k\rho)\right) & \\
\mathbf{e}_{z}\frac{ik}{k_{1z}^{+}}\left\{ \begin{array}{c}
-\sin\left(k_{1z}^{+}z\right)\\
\cos\left(k_{1z}^{+}d\right)
\end{array}\right\} i\cos\left(\varphi_{\boldsymbol{\rho}}\right)J_{1}(k\rho)+\left(\mathbf{e}_{x}\left[\cos^{2}\left(\varphi_{\boldsymbol{\rho}}\right)J_{0}(k\text{\ensuremath{\rho}})-\frac{\cos\left(2\varphi_{\boldsymbol{\rho}}\right)J_{1}(k\rho)}{k\text{\ensuremath{\rho}}}\right]-\mathbf{e}_{y}\frac{\sin\left(2\varphi_{\boldsymbol{\rho}}\right)}{2}J_{2}(k\rho)\right)\left\{ \begin{array}{c}
\cos\left(k_{1z}^{+}d\right)\\
\sin\left(k_{1z}^{-}d\right)
\end{array}\right\} ,& \\
e^{ik_{2z}\left(z-d\right)}\left\{ \begin{array}{c}
\cos\left(k_{1z}^{+}d\right)\\
\sin\left(k_{1z}^{-}d\right)
\end{array}\right\} \left(-\mathbf{e}_{z}\frac{k}{k_{2z}}i\cos\left(\varphi_{\boldsymbol{\rho}}\right)J_{1}(k\rho)+\mathbf{e}_{x}\left[\cos^{2}\left(\varphi_{\boldsymbol{\rho}}\right)J_{0}(k\text{\ensuremath{\rho}})-\frac{\cos\left(2\varphi_{\boldsymbol{\rho}}\right)J_{1}(k\rho)}{k\text{\ensuremath{\rho}}}\right]-\mathbf{e}_{y}\frac{\sin\left(2\varphi_{\boldsymbol{\rho}}\right)}{2}J_{2}(k\rho)\right) & 
\end{array}\right.
\label{eq:dipole_parallel_TM_after_integration}
\end{gather}
where the upper, middle and bottom lines are for Regions I, II and III respectively. $\varphi_{\boldsymbol{\rho}}$ denotes the angle of $\boldsymbol{\rho}$ with respect to  $\mathbf{e}_{x}.$

For $\rho=0$ we obtain
\begin{equation}
\int\frac{\left\langle \mathbf{\tilde{E}_{k\,\mathrm{TM}}^{\pm}}|\mathbf{E}_{0}\right\rangle }{L_{x}L_{y}}\mathbf{E}_{\mathbf{k}}^{\pm}d\varphi_\mathbf{k}=\frac{4\pi ps_{k}^{\pm}}{\epsilon_{2}}e^{ik_{2z}\left(z_{0}-d\right)}\left\{ \begin{array}{c}
\cos\left(k_{1z}^{+}d\right)\\
\sin\left(k_{1z}^{-}d\right)
\end{array}\right\} \pi\left(B_{k}^{\pm}\right)^{2}\left\{ \begin{array}{cc}
e^{-ik_{2z}\left(z+d\right)}\left\{ \begin{array}{c}
\cos\left(k_{1z}^{+}d\right)\\
-\sin\left(k_{1z}^{-}d\right)
\end{array}\right\} \mathbf{e}_{x} & \mathbf{r}\in\mathbf{\mathrm{I}}\\
\left\{ \begin{array}{c}
\cos\left(k_{1z}^{+}d\right)\\
\sin\left(k_{1z}^{-}d\right)
\end{array}\right\} \mathbf{e}_{x} & \mathbf{r}\in\mathbf{\mathrm{II}}\\
e^{ik_{2z}\left(z-d\right)}\left\{ \begin{array}{c}
\cos\left(k_{1z}^{+}d\right)\\
\sin\left(k_{1z}^{-}d\right)
\end{array}\right\} \mathbf{e}_{x} & \mathbf{r}\in\mathbf{\mathrm{III}}
\end{array}\right..
\label{eq:dipole_parallel_TM_after_integration_rho_zero}
\end{equation}
\end{widetext}

\subsubsection{Contribution of the TE modes}
By substituting $\mathbf{e}_{\perp}=\sin\left(\varphi_{\mathbf{k}}\right)\mathbf{e}_{x}-\cos\left(\varphi_{\mathbf{k}}\right)\mathbf{e}_{y}$ we arrive at the following expression for the TE eigenfunctions

\begin{widetext}
$$\mathbf{E}_{\mathbf{k}}^{\pm}=e^{i\mathbf{k}\cdot\rho}\left\{ \begin{array}{cc}
\left(\sin\left(\varphi_{\mathbf{k}}\right)\mathbf{e}_{x}-\cos\left(\varphi_{\mathbf{k}}\right)\mathbf{e}_{y}\right)B_{\perp}^{\pm}\left\{ \begin{array}{c}
\cos\left(k_{1z}^{+}d\right)\\
-\sin\left(k_{1z}^{-}d\right)
\end{array}\right\} e^{-ik_{2z}\left(z+d\right)} & \mathbf{r}\in\mathbf{\mathrm{I}}\\
\left(\sin\left(\varphi_{\mathbf{k}}\right)\mathbf{e}_{x}-\cos\left(\varphi_{\mathbf{k}}\right)\mathbf{e}_{y}\right)B_{\perp}^{\pm}\left\{ \begin{array}{c}
\cos\left(k_{1z}^{+}z\right)\\
\sin\left(k_{1z}^{-}z\right)
\end{array}\right\}  & \mathbf{r}\in\mathbf{\mathrm{II}}\\
\left(\sin\left(\varphi_{\mathbf{k}}\right)\mathbf{e}_{x}-\cos\left(\varphi_{\mathbf{k}}\right)\mathbf{e}_{y}\right)B_{\perp}^{\pm}\left\{ \begin{array}{c}
\cos\left(k_{1z}^{+}z\right)\\
\sin\left(k_{1z}^{-}z\right)
\end{array}\right\} e^{ik_{2z}\left(z-d\right)} & \mathbf{r}\in\mathbf{\mathrm{III}}
\end{array}\right..$$

The scalar product $\left\langle \mathbf{\tilde{E}_{k\,\mathrm{TE}}^{\pm}}|\mathbf{E}_{0}\right\rangle$ can be written as
$$\frac{\left\langle \mathbf{\tilde{E}_{k\,\mathrm{TE}}^{\pm}}|\mathbf{E}_{0}\right\rangle }{L_{x}L_{y}}=\frac{4\pi ps_{k}^{\pm}}{\epsilon_{2}}B_{\perp}^{+}\left\{ \begin{array}{c}
\cos\left(k_{1z}^{+}d\right)\\
\sin\left(k_{1z}^{-}d\right)
\end{array}\right\} e^{ik_{2z}\left(z_{0}-d\right)}\sin\left(\varphi_{\mathbf{k}}\right). $$
Integrating $\frac{\left\langle \mathbf{\tilde{E}_{k\,\mathrm{TE}}^{\pm}}|\mathbf{E}_{0}\right\rangle }{L_{x}L_{y}}\mathbf{E}_{\mathbf{k}}^{\pm}$ with respect to $\varphi_\mathbf{k}$ we obtain
\[
\int\frac{\left\langle \mathbf{\tilde{E}_{k\,\mathrm{TE}}^{\pm}}|\mathbf{E}_{0}\right\rangle }{L_{x}L_{y}}\mathbf{E}_{\mathbf{k}}^{\pm}d\varphi_{\mathbf{k}}=\frac{4\pi ps_{k}^{\pm}}{\epsilon_{2}}\left(B_{\perp}^{\pm}\right)^{2}\cos\left(k_{1z}^{\pm}d\right)e^{ik_{2z}\left(z_{0}-d\right)} \times
\]
\begin{equation}
2\pi\left\{ \begin{array}{cc}
\left[\left(\sin^{2}\left(\varphi_{\boldsymbol{\rho}}\right)J_{0}\left(k\rho\right)+\frac{\cos\left(2\varphi_{\boldsymbol{\rho}}\right)J_{1}(k\rho)}{k\text{\ensuremath{\rho}}}\right)\mathbf{e}_{x}+\frac{1}{2}\sin\left(2\varphi_{\boldsymbol{\rho}}\right)J_{2}\left(k\rho\right)\mathbf{e}_{y}\right]\left\{ \begin{array}{c}
\cos\left(k_{1z}^{+}d\right)\\
-\sin\left(k_{1z}^{-}d\right)
\end{array}\right\} e^{-ik_{2z}\left(z+d\right)} & \mathbf{r}\in\mathbf{\mathrm{I}}\\
\left[\left(\sin^{2}\left(\varphi_{\boldsymbol{\rho}}\right)J_{0}\left(k\rho\right)+\frac{\cos\left(2\varphi_{\boldsymbol{\rho}}\right)J_{1}(k\rho)}{k\text{\ensuremath{\rho}}}\right)\mathbf{e}_{x}+\frac{1}{2}\sin\left(2\varphi_{\boldsymbol{\rho}}\right)J_{2}\left(k\rho\right)\mathbf{e}_{y}\right]\left\{ \begin{array}{c}
\cos\left(k_{1z}^{+}z\right)\\
\sin\left(k_{1z}^{-}z\right)
\end{array}\right\}  & \mathbf{r}\in\mathbf{\mathrm{II}}\\
\left[\left(\sin^{2}\left(\varphi_{\boldsymbol{\rho}}\right)J_{0}\left(k\rho\right)+\frac{\cos\left(2\varphi_{\boldsymbol{\rho}}\right)J_{1}(k\rho)}{k\text{\ensuremath{\rho}}}\right)\mathbf{e}_{x}+\frac{1}{2}\sin\left(2\varphi_{\boldsymbol{\rho}}\right)J_{2}\left(k\rho\right)\mathbf{e}_{y}\right]\left\{ \begin{array}{c}
\cos\left(k_{1z}^{+}d\right)\\
\sin\left(k_{1z}^{-}d\right)
\end{array}\right\} e^{ik_{2z}\left(z-d\right)} & \mathbf{r}\in\mathbf{\mathrm{III}}
\end{array}\right..
\label{eq:dipole_parallel_TE_after_integration}
\end{equation}

For $\rho=0$ we obtain

\begin{equation}
 \int\frac{\left\langle \mathbf{\tilde{E}_{k\,\mathrm{TE}}^{\pm}}|\mathbf{E}_{0}\right\rangle }{L_{x}L_{y}}\mathbf{E}_{\mathbf{k}}^{\pm}d\varphi_{\mathbf{k}}=\frac{4\pi ps_{k}^{\pm}}{\epsilon_{2}}\left(B_{\perp}^{\pm}\right)^{2}\cos\left(k_{1z}^{\pm}d\right)e^{ik_{2z}\left(z_{0}-d\right)}\pi\left\{ \begin{array}{cc}
\mathbf{e}_{x}\left\{ \begin{array}{c}
\cos\left(k_{1z}^{+}d\right)\\
-\sin\left(k_{1z}^{-}d\right)
\end{array}\right\} e^{-ik_{2z}\left(z+d\right)} & \mathbf{r}\in\mathbf{\mathrm{I}}\\
\mathbf{e}_{x}\left\{ \begin{array}{c}
\cos\left(k_{1z}^{+}z\right)\\
\sin\left(k_{1z}^{-}z\right)
\end{array}\right\}  & \mathbf{r}\in\mathbf{\mathrm{II}}\\
\mathbf{e}_{x}\left\{ \begin{array}{c}
\cos\left(k_{1z}^{+}d\right)\\
\sin\left(k_{1z}^{-}d\right)
\end{array}\right\} e^{ik_{2z}\left(z-d\right)} & \mathbf{r}\in\mathbf{\mathrm{III}}
\end{array}\right..
\label{eq:dipole_parallel_TE_after_integration_rho_zero}\end{equation}
\end{widetext}

When $k\thickapprox0$ for the second odd TE mode in which $\epsilon_{1}^{-} \thickapprox0$ we get that $k_{1z}^{-}=\sqrt{k_{0}^{2}\epsilon_{1}^{-}-k^{2}} \thickapprox0$ and $\langle\tilde{{\bf E}}_{{\bf k}}^{-}|{\bf E}_{{\bf k}}^{-}\rangle \thickapprox0.$
To avoid numerical inaccuracies we approximate  $\langle\tilde{{\bf E}}_{{\bf k}}^{-}|{\bf E}_{{\bf k}}^{-}\rangle$ for $k_{1z}^{-}\eqsim0\,$ as follows
\begin{align}
\frac{\langle\tilde{{\bf E}}_{{\bf k}}^{-}|{\bf E}_{{\bf k}}^{-}\rangle}{L_{x}L_{y}\left(B_{\perp}^{-}\right)^{2}}=d-\sin\left(2k_{1z}^{-}d\right)/2k_{1z}^{-}\nonumber \\
\eqsim d-\frac{\left(2k_{1z}^{-}d\right)-\frac{\left(2k_{1z}^{-}d\right)^{3}}{3}}{2k_{1z}^{-}}=\frac{4}{3}\left(k_{1z}^{-}\right)^{2}d^{3}.\nonumber
\end{align}

Eq. (\ref{Eexpansion}) now reads
\begin{align}
\label{Eexpansion4}
&|{\bf E}\rangle-|{\bf E}_{0}\rangle \nonumber \\
&=\sum_{\mathrm{TM,TE}}\sum_{+,-}\int\frac{dk}{\left(2\pi\right)^{2}}\frac{s_{\mathbf{k}}}{s-s_{\mathbf{k}}}\frac{\left(\int\langle\tilde{{\bf E}}_{\mathbf{k}}|{\bf E}_{0}\rangle|{\bf E}_{\mathbf{k}}\rangle d\varphi_{\mathbf{k}}\right)}{\langle\tilde{{\bf E}}_{\mathbf{k}}|{\bf E}_{\mathbf{k}}\rangle}k  \nonumber \\
&\equiv\sum_{\mathrm{TM,TE}}\sum_{+,-}\int dk\mathbf{F}\left(\mathbf{r},k\right),
 \end{align}
where $\int\langle\tilde{{\bf E}}_{\mathbf{k}}|{\bf E}_{0}\rangle|{\bf E}_{\mathbf{k}}\rangle d\varphi_{\mathbf{k}}$ is given by Eqs. (\ref{eq:dipole_parallel_TM_after_integration}),(\ref{eq:dipole_parallel_TM_after_integration_rho_zero}),(\ref{eq:dipole_parallel_TE_after_integration}) and (\ref{eq:dipole_parallel_TE_after_integration_rho_zero}) and the expressions for $s_k$ and $\langle\tilde{{\bf E}}_{\mathbf{k}}|{\bf E}_{\mathbf{k}}\rangle$ are given in Section III.

\subsubsection{Calculation of the integrands as functions of $k$}

\begin{figure}[t]
   \begin{center}
   \begin{tabular}{c}
   \includegraphics[width=9cm]{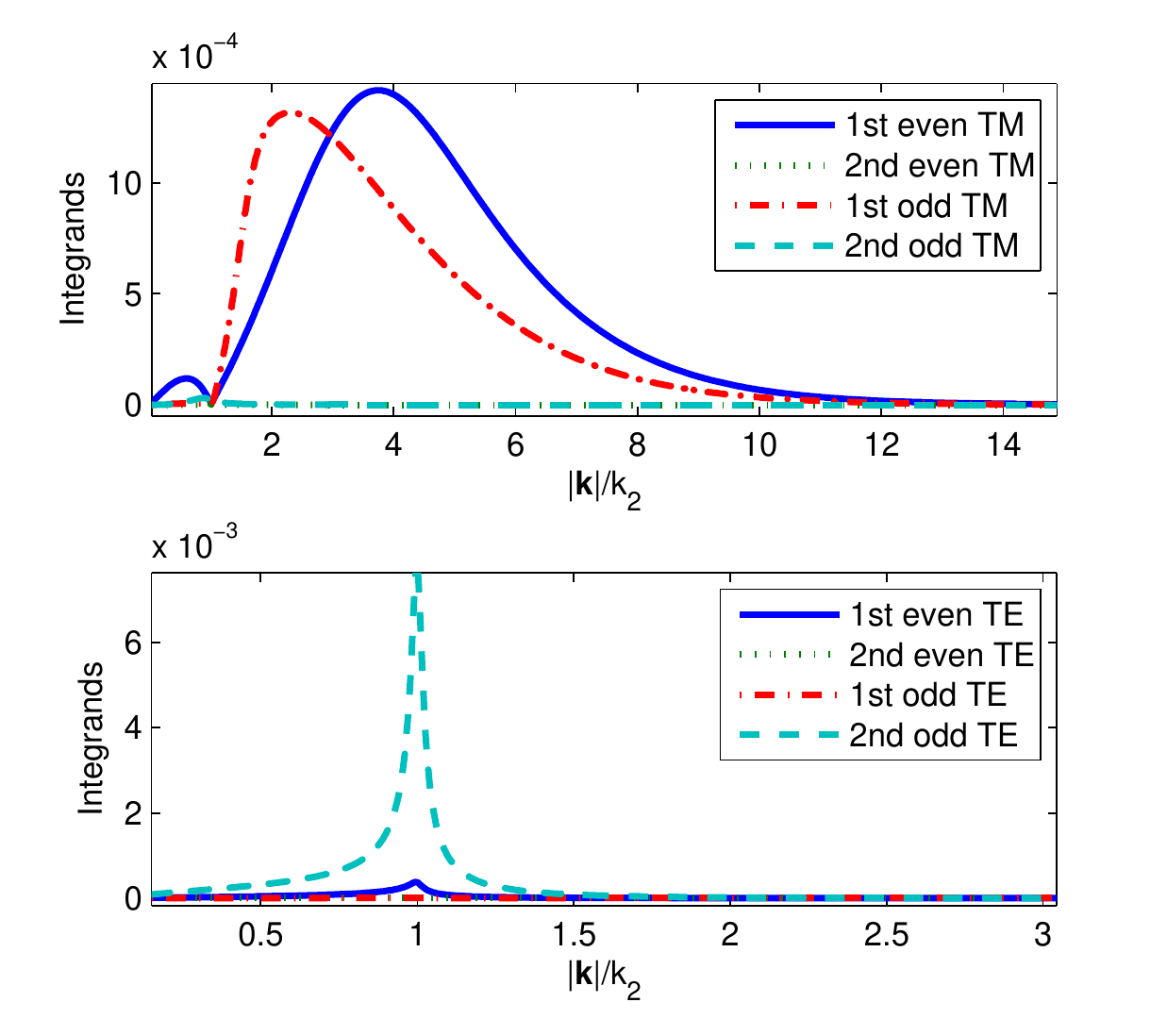}
   \end{tabular}
   \end{center}
   \caption[]
   { \label{fig:integrands_parallel} 
Integrands of Eq.  (\ref{Eexpansion4}) for the first two even and odd TM and TE modes at $\rho=0,z=-d.$}
   \end{figure}

We calculated the integrands in Eq.\ (\ref{Eexpansion4}) for the first two even and odd TM and TE modes for the coordinates $z=-d,\rho=0.$  In Fig. \ref{fig:integrands_parallel} we present the absolute value of the integrands as functions of $\left|\mathbf{k}\right|$. It can be seen that the modes with the dominant contribution to the expansion are the first even and odd TM modes and the second odd TE mode. While the contribution of the first even and odd TM modes originates from $s_k/(s-s_k)$ since $\epsilon_{1k}\thickapprox\epsilon_1$ for $k>k_2,$ the contribution of the second odd TE mode originates from $s_k$ which appears when there are current sources since $\epsilon_{1k}\thickapprox\epsilon_2$ for $k\thickapprox k_2.$

\subsubsection{Calculation of the electric field}

\begin{figure}[t]
   \begin{center}
   \begin{tabular}{c}
   \includegraphics[height=6cm]{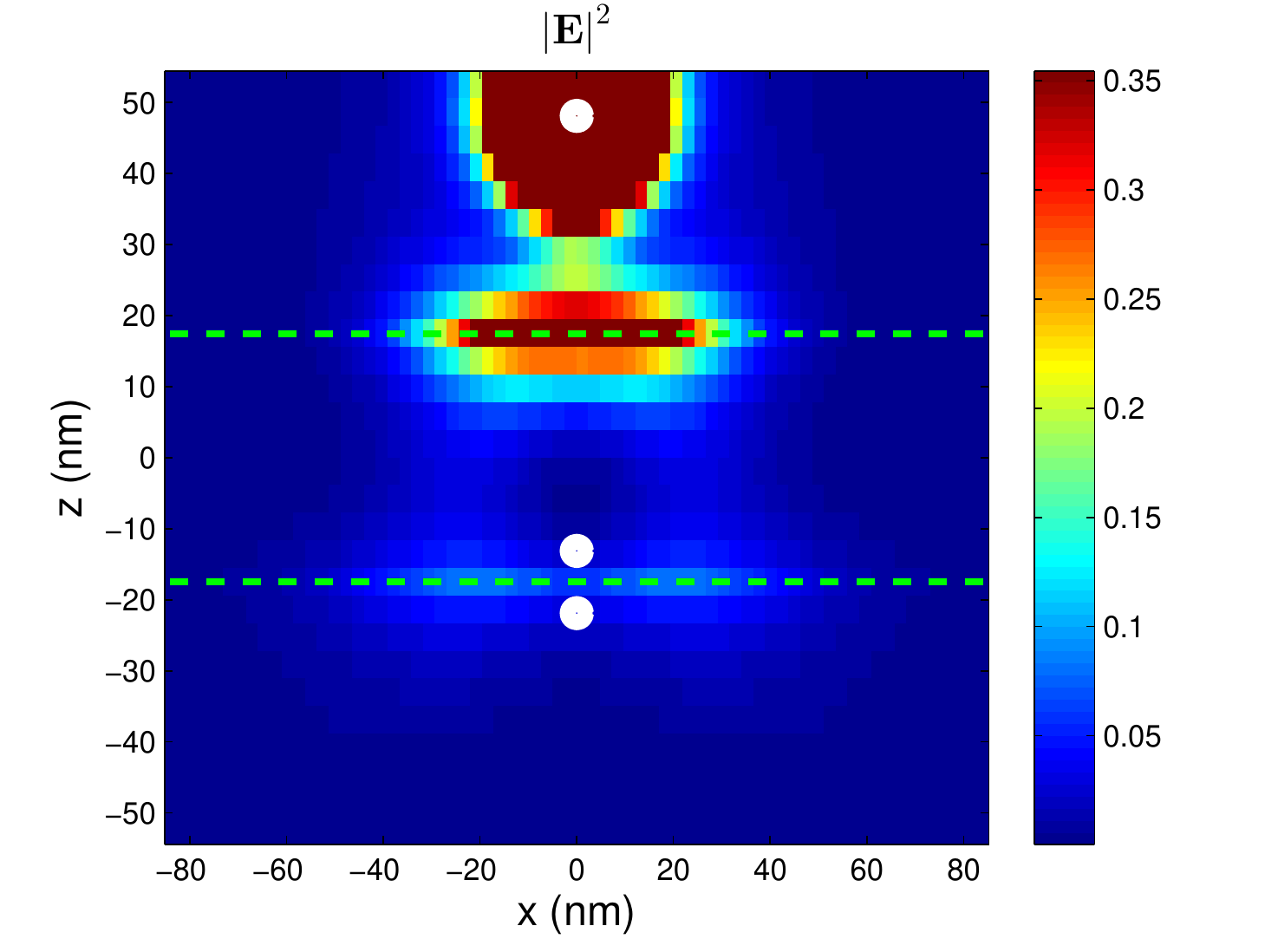}
   \end{tabular}
   \end{center}
   \caption[]
   { \label{fig:intensity_parallel} 
$\left|\mathbf{E}\right|^{2}$ in the three regions for a dipole directed along $x$ located at $\rho=0, z=d+7\cdot\left(2d\right)/8$  and \mbox{$\epsilon_1=-2.55+0.24i,$} \mbox{$\epsilon_2=2.57+0.09i.$}}
  \end{figure}

\begin{figure}[t]
   \begin{center}
   \begin{tabular}{c}
   \includegraphics[height=6cm]{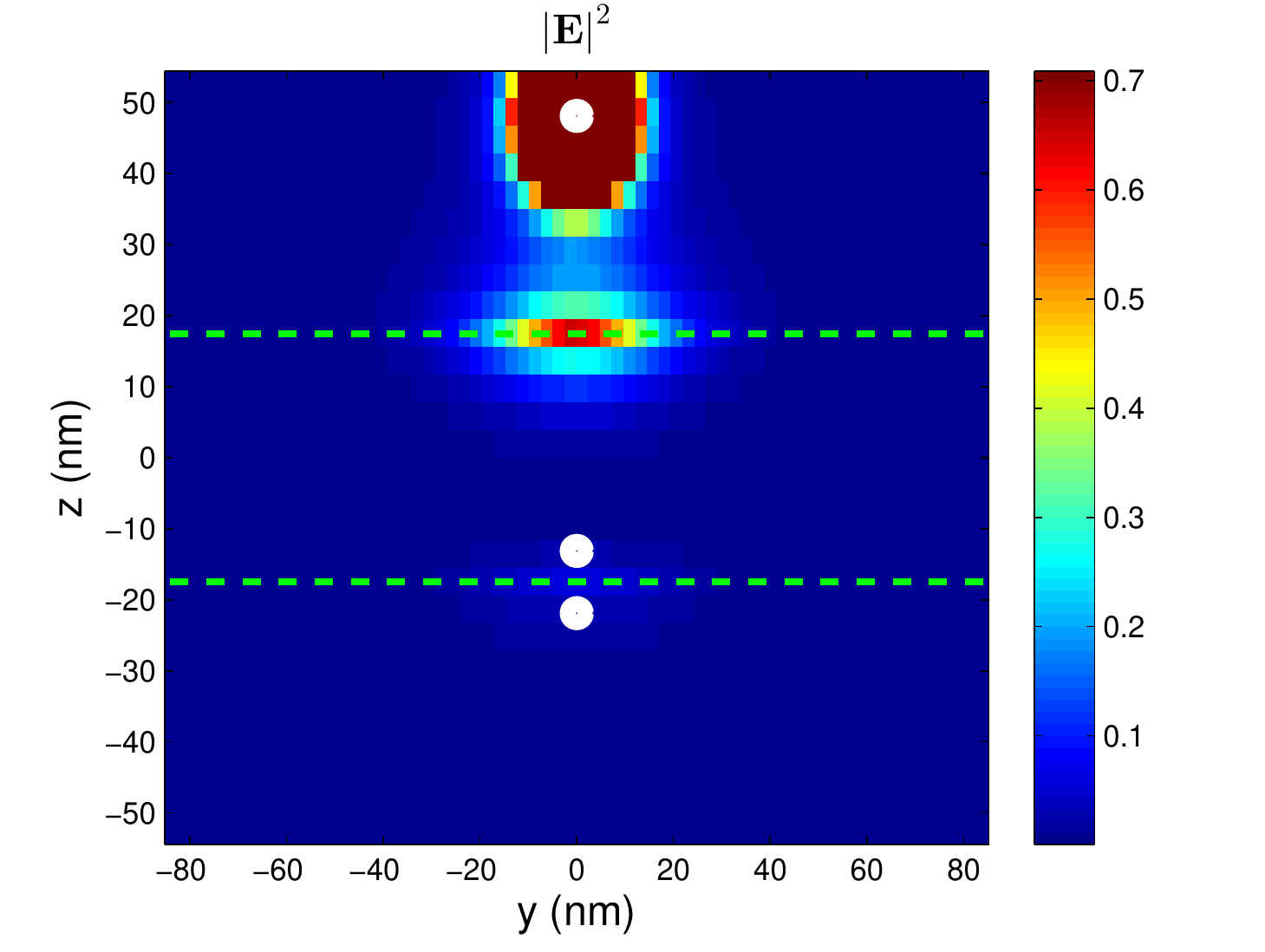}
   \end{tabular}
   \end{center}
   \caption[]
   { \label{fig:intensity_parallel_2} 
$\left|\mathbf{E}\right|^{2}$ in the three regions for a  dipole directed along $x$ located at $\rho=0, z=d+7\cdot\left(2d\right)/8$  and \mbox{$\epsilon_1=-2.55+0.24i,$} \mbox{$\epsilon_2=2.57+0.09i.$}}
  \end{figure}

We calculated the electric field  in the three regions  by numerically integrating Eq. (\ref{Eexpansion4}) with respect to $k.$ In Fig. \ref{fig:intensity_parallel} we present the intensity at the $y=0$ plane for a dipole directed along $x$ axis located at $z=d+7\cdot\left(2d\right)
 /8$ and permittivity values of \mbox{$\epsilon_1=-2.55+0.24i, \epsilon_2=2.57+0.09i$}.
 It can be seen that the intensity peaks at the top interface at $x=0$ and at the bottom interface there are two peaks at $x=-23$ nm and $x=23$nm. In Fig. \ref{fig:intensity_parallel_2} we present the intensity for the dipole at the $x=0$ plane. It can be seen that  intensity peaks at the top and bottom interfaces at $y=0$ and that the horizontal width of the intensity is smaller compared to the previous case.

\begin{figure}[t]
   \begin{center}
   \begin{tabular}{c}
   \includegraphics[width=7cm]{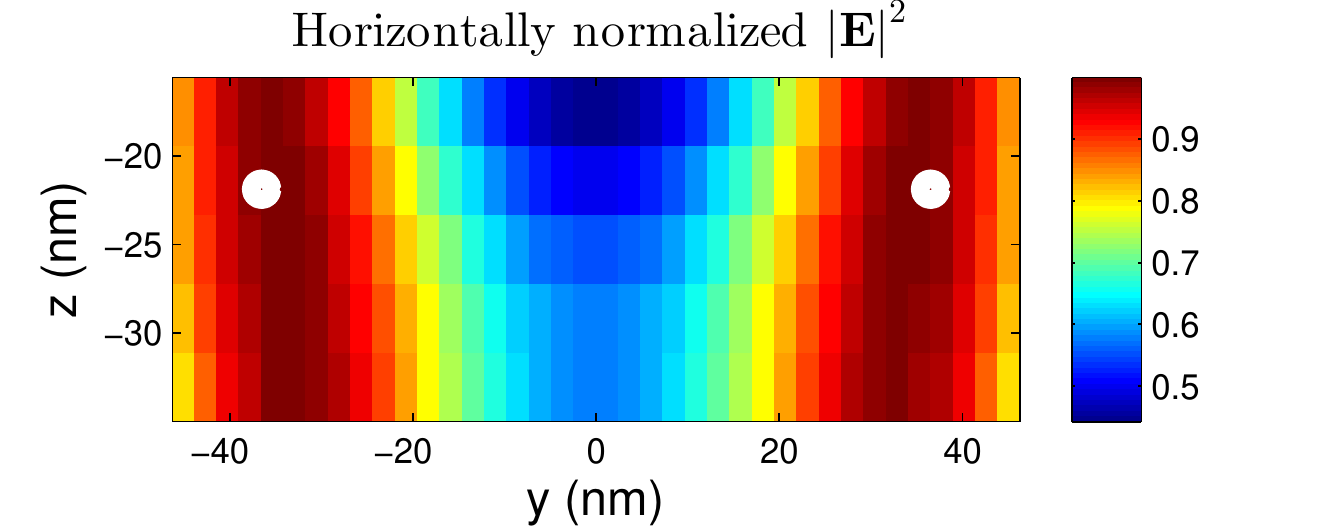}
   \end{tabular}
   \end{center}
   \caption[]
   { \label{fig:resolution_parallel_y} 
Horizontally normalized $\left|\mathbf{E}\right|^{2}$ in Region I for two oscillating dipoles shifted in the $y$ axis located at \mbox{$z=d+7\cdot\left(2d\right)/8$} and $\epsilon_1=-2.55+0.24i, \epsilon_2=2.57+0.09i.$}
  \end{figure}

\begin{figure}[t]
   \begin{center}
   \begin{tabular}{c}
   \includegraphics[width=7cm]{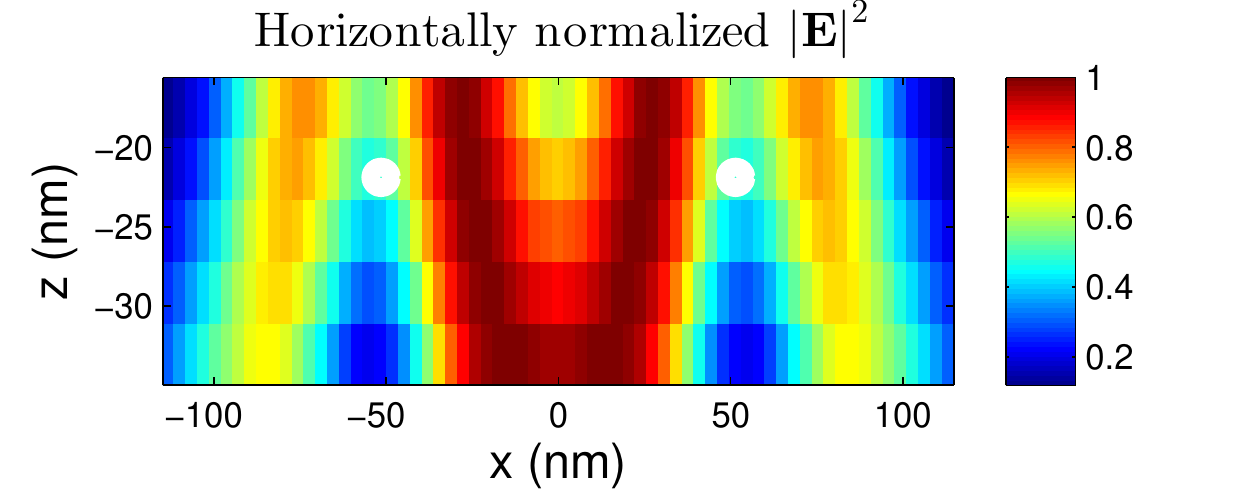}
   \end{tabular}
   \end{center}
   \caption[]
   { \label{fig:resolution_parallel_x} 
Horizontally normalized $\left|\mathbf{E}\right|^{2}$ in Region I for two oscillating dipoles shifted in the $x$ axis located at \mbox{$z=d+7\cdot\left(2d\right)/8$} and $\epsilon_1=-2.55+0.24i, \epsilon_2=2.57+0.09i.$}
  \end{figure}

 In Fig. \ref{fig:resolution_parallel_y} we present the horizontally normalized $\left|\mathbf{E}\right|^{2}$ in Reg. I for two dipole objects shifted in the $y$ axis. The white circles denote the images expected according to geometric optics. It can be seen that the optimal separation between the images is at the interface. In Fig. \ref{fig:resolution_parallel_x} we present the horizontally normalized $\left|\mathbf{E}\right|^{2}$ in Reg. I for two dipole objects shifted in the $x$ axis. Since each image is approximately composed of a sum two Gaussians we regarded the separation between the images as the separation between the two internal Gaussians (higher intensity due to constructive interference). Here too, the optimal separation between the images is at the interface.

\subsubsection*{Verification of the results}
To verify our results we checked the continuity of the physical $D_z$ at the interfaces. This continuity is not trivially satisfied as the continuity of $D_z$ of each eigenmode is satisfied for the eigenvalue $\epsilon_{1\mathbf{k}}$ and not for the physical value of $\epsilon_1$. In fact each term in Eq. (\ref{Eexpansion}) usually violates the continuity of $D_z$ at the interfaces for the physical permittivity values. We calculated $D_z$ throughout the interfaces and it was found to be continuous to a very high precision for the perpendicular and parallel dipole calculations. 

\section{Discussion}
\label{discussion}
We presented an exact calculation of the local electric field ${\bf E}({\bf r})$ for a setup of an $\epsilon_1$ slab in an $\epsilon_2$ medium and  a
time dependent electric point dipole ${\bf p}e^{-i\omega t}$ situated in the medium and directed parallel and perpendicular to the slab. For this purpose we first reformulated the differential equation which follows from Maxwell's equations as an integro-differential equation and expressed ${\bf E}({\bf r})$ in terms of the eigenfunctions of the setup. We constructed all the TE and TM modes for the setup using its symmetry properties. We then simplified the calculation of $\langle\tilde{\bf E}_n|{\bf E}_0\rangle$ for external current sources in order to enable that calculation to be performed analytically. We calculated the eigenvalues of the even and odd TM and TE modes as functions of $\left|\mathbf{k}\right|.$ Finally, we calculated $\left|\mathbf{E}\right|^{2}$ and the horizontal resolution for permittivity values which match the PMMA-silver-photoresist experiment \cite{Zhang} and for a setup in which $\Delta  s$ is divided by 1000.

The set of eigenvalues $\epsilon_{1\mathbf{k}}$ are non-physical values which are determined by the values of $\epsilon_2,\lambda$ and $d$. When the physical value of $\epsilon_{1}\left(\omega\right)$ of the slab becomes closer to $\epsilon_{1\mathbf{k}},$ the incoming EM wave with the same $\left|\mathbf{k}\right|$ will experience amplification. Since the eigenvalues $\epsilon_{1\mathbf{k}}$ sometimes tend to $-\epsilon_2$ when $\left|\mathbf{k}\right| \rightarrow \infty$, a slab with $\epsilon_{1}\thickapprox-\epsilon_{2}$ will amplify the evanescent waves, resulting in enhanced resolution. When $\epsilon_1$ further approaches $-\epsilon_2,$ modes with higher $\left|\mathbf{k}\right|$ will also be employed in the imaging, resulting in further enhanced resolution. The optimal imaging, as in our quasistatic analysis \cite{BergPRA2014,FarhiBergVeselagoPRA2014}, was found to be not at the geometric optics foci but at the interface between the slab and Region I.
 In addition, when there are current sources an additional $s_k$ factor is introduced, resulting in another singularity when $\epsilon_{1k}\thickapprox \epsilon_2.$
  Interestingly, the second odd TM and TE modes in the range where $\left|\mathbf{k}\right|\thickapprox 0$ have $\epsilon_{1\mathbf{k}}^-\thickapprox 0$. The second odd TE mode in the range $\left|\mathbf{k}\right|<k_2$ has $\epsilon_{1\mathbf{k}}^-$  which are close to real. These ranges of $\epsilon_{1\mathbf{k}}$ are not far from $\epsilon_1$ values which are realizable in experiments and may have implications in optical devices where amplification of optical signals is important. In addition, since for the second odd TE mode  $\epsilon_{1k}\thickapprox\epsilon_2$ for $k\thickapprox k_2$ there is an enhancement of the electric field due to the additional $s_k$ factor which is introduced when there are current sources.

The formalism enables to calculate the electric field of oscillating current sources in a simple manner, avoiding the complex calculation of the scattering of the electric field emanating from these sources.  Since current sources are used to represent polarized media and objects in imaging, the formalism can find use in many applications.

The propagating and evanescent eigenstates are related to the incoming propagating and evanescent waves, respectively, through $\left\langle \mathbf{E_{-k}^{\mp}}|\mathbf{E}_{0}\right\rangle$. Since \mbox{$\ensuremath{\langle\mathbf{E}_{-\mathbf{k}_{1}}^{\mp}|{\bf E}_{\mathbf{k}_{2}}^{\mp}\rangle}\varpropto\delta^{2}\left(\mathbf{k}_{2}-\mathbf{k}_{1}\right),$} when the the incoming EM waves include waves with a given $\mathbf{k}$ as the 2D vector, the eigenmodes with the same $\mathbf{k}$ will contribute to the expansion. Thus, if the object is far from the slab and the evanescent waves reach the slab with low amplitude, the scalar product $\left\langle \mathbf{E_{-k}^{\mp}}|\mathbf{E}_{0\,\mathbf{k}}\right\rangle$ (where $\left|\mathbf{E}_{0\,\mathbf{k}}\right\rangle$ denotes the $\mathbf{k}$ component of the source) will be small and the evanescent modes will have a low contribution to the expansion of the electric field.

\acknowledgments     

Y. Sivan is acknowledged for useful comments.
This work was supported, in part, by a grant from MAFAT.


\appendix
\section{The left and right eigenstates of a symmetric operator as a bi-orthogonal
basis in Hilbert space}
\label{biorthogonal}
This Appendix is based upon material covered in Section II of Ref.\ \onlinecite{BergStroudPRB80}.

For any state $|\psi\rangle$ in Hilbert space we define the ``dual state'' 
$|\tilde{\psi}\rangle$ by citing the following relation for its wave function representation
$\langle{\bf r}|\tilde{\psi}\rangle$
\be
\langle{\bf r}|\tilde{\psi}\rangle\equiv\langle{\bf r}|\psi\rangle^*
 =\langle\psi|{\bf r}\rangle.
\label{DualStateDef}
\ee
An operator $\hat\Gamma$ will be called  symmetric  if it satisfies
$$
\langle\tilde\phi|\hat\Gamma|\psi\rangle=\langle\tilde\psi|\hat\Gamma|\phi\rangle
$$
for any two states $|\phi\rangle$, $|\psi\rangle$ in Hilbert space.
Using the wave function representation for these states we can write their scalar product
$\langle\tilde\phi|\psi\rangle$ in the following explicit form as an intergral over space
\be
\langle\tilde\phi|\psi\rangle=\int d^3r\langle\tilde\phi|{\bf r}\rangle
 \langle{\bf r}|\psi\rangle=\int d^3r \langle{\bf r}|\phi\rangle \langle{\bf r}|\psi\rangle
  =\langle\tilde\psi|\phi\rangle.
\label{ScalarProd}
\ee

If $|\psi_n\rangle$ is a right eigenstate of the symmetric operator
$\hat\Gamma$ 
$$\hat\Gamma|\psi_n\rangle=s_n|\psi_n\rangle$$
then $\langle\tilde\psi_n|$ is a left eigenstate of $\hat\Gamma$
with same eigenvalue $s_n$ since the following holds for any state
$|\psi\rangle$ in Hilbert space:
\begin{eqnarray*}
\langle\tilde{\psi}_n|\hat\Gamma|\psi\rangle&=&\langle\tilde{\psi}|\hat\Gamma|\psi_n\rangle
 =s_n\langle\tilde{\psi}|\psi_n\rangle=s_n\langle\tilde{\psi}_n|\psi\rangle,
\end{eqnarray*}
therefore
\be
\langle\tilde{\psi}_n|\hat\Gamma=s_n\langle\tilde{\psi}_n|.
\label{DualEigenstate}
\ee

By considering a pair of right and left eigenstates we get that
$$
\langle\tilde{\psi}_m|\hat\Gamma|\psi_n\rangle=s_m\langle\tilde{\psi}_m|\psi_n\rangle=
 s_n\langle\tilde{\psi}_m|\psi_n\rangle.
$$
Thus, if $s_m\neq s_n$ then these two states must be mutually orthogonal, i.e.,
$\langle\tilde{\psi}_m|\psi_n\rangle=\langle\tilde{\psi}_n|\psi_m\rangle=0$.
Such a set of states is called a bi-orthogonal set.
From Eq.\ (\ref{ScalarProd}) it follows that the scalar product of any state
$|\psi\rangle$ and its dual $|\tilde\psi\rangle$ becomes
\be
\langle\tilde\psi|\psi\rangle=\int d^3r \langle{\bf r}|\psi\rangle^2.
\label{BiorthogonalNorm}
\ee
Because the integrand is not necessarily positive nor even real, this integral
could possibly vanish. In order for the set of right eigenstates of $\hat\Gamma$
to be a complete set in Hilbert space, i.e., a basis,  the scalar product of any
eigenstate $|\psi_n\rangle$ and its dual must be nonzero. This needs to be verified
for all the eigenstates. If this requirement is satisfied then the unit operator can
be written as
\be
\openone=\sum_n \frac{|\psi_n\rangle\langle\tilde\psi_n|}{\langle\tilde{\psi}_n|\psi_n\rangle}
\label{UnityOp}
\ee
and the state $|\psi\rangle$ can be expanded in a series of  the right eigenstates  $|\psi_n\rangle$
\be
|\psi\rangle=\sum_n |\psi_n\rangle\frac{\langle\tilde{\psi}_n|\psi\rangle}
 {\langle\tilde{\psi}_n|\psi_n\rangle}.
\label{psiExpansion}
\ee
These eigenstates are called a ``bi-orthogonal basis'' of Hibert space.

A complication arises when eigenstates of $\hat\Gamma$ are degenerate due to the
existence of symmetry operators. Those are one or more Hermitian or unitary operators
$\hat P$ that commute with $\hat\Gamma$. In that case we often like to select eigenstates of
$\hat\Gamma$ that are also eigenstates of $\hat P$. A difficulty occurs when the complex
conjugation that leads to the dual eigenstate of $\hat\Gamma$ results in a state which
is not an eigenstate of $\hat P$ or is an eigenstate of $\hat P$ with a different eigenvalue.
Such a situation occurs in the case of a spherical inclusion and also in the case of a flat slabs
microstructure. The first case was discussed in Ref.\ \onlinecite{BergStroudPRB80} while
the second case is discussed in Section \ref{FlatSlabs} of the current article.

\section{Quasistatic results}
\subsection{Flat-slab modes}

By taking the quasi-static limit $k_{0}\rightarrow0$
  we obtain the following results for the TM modes which are associated with electro-statics
\begin{gather}
\mathbf{E}_{\mathbf{k}}^{+}=e^{i\mathbf{k}\cdot\boldsymbol{\rho}}\left\{ \begin{array}{cc}
e^{kz}A_{k}^{+}\left(-i\mathbf{e}_{z}+\mathbf{e}_{\mathbf{k}}\right) & \mathbf{r}\in\mathbf{\mathrm{I}}\\
B_{k}^{+}\left(-\mathbf{e}_{z}i\sinh\left(kz\right)+\mathbf{e}_{\mathbf{k}}\cosh\left(kz\right)\right) & \mathbf{r}\in\mathbf{\mathrm{II}}\\
e^{-kz}A_{k}^{+}\left(i\mathbf{e}_{z}+\mathbf{e}_{\mathbf{k}}\right) & \mathbf{r}\in\mathbf{\mathrm{III}}
\end{array}\right.,\\
\mathbf{E}_{\mathbf{k}}^{-}=e^{i\mathbf{k}\cdot\boldsymbol{\rho}}\left\{ \begin{array}{cc}
e^{kz}A_{k}^{-}\left(-i\mathbf{e}_{z}+\mathbf{e}_{\mathbf{k}}\right) & \mathbf{r}\in\mathbf{\mathrm{I}}\\
B_{k}^{-}\left(\mathbf{e}_{z}\cosh\left(kz\right)+\mathbf{e}_{\mathbf{k}}i\sinh\left(kz\right)\right) & \mathbf{r}\in\mathbf{\mathrm{II}}\\
e^{-kz}A_{k}^{-}\left(-i\mathbf{e}_{z}-\mathbf{e}_{\mathbf{k}}\right) & \mathbf{r}\in\mathbf{\mathrm{III}}
\end{array}\right.,\\
\frac{\langle\tilde{{\bf E}}_{{\bf k}}^{\pm}|{\bf E}_{{\bf k}}^{\pm}\rangle}{L_{x}L_{y}}=\mp\frac{\left(B_{k}^{\pm}\right)^{2}}{k}\sinh\left(2kd\right).
  \end{gather}

\subsection{Results for a point dipole in Reg. III directed along z}
We calculated $\left\langle \mathbf{\tilde{{\bf E}}_{{\bf k}}^{\pm}}|\mathbf{E}_{0}\right\rangle$  in the quasi-static limit and obtained

$$\frac{\left\langle \mathbf{\tilde{E}_{k\,\mathrm{TM}}^{\pm}}|\mathbf{E}_{0}\right\rangle }{L_{x}L_{y}}=B_{k}^{\pm}\left\{ \begin{array}{c}
-i\cosh(dk)\\
\sinh(dk)
\end{array}\right\} \frac{4\pi ps_{k}^{\pm}e^{-k(z_{0}-d)}}{\epsilon_{2}}. $$
\begin{widetext}
We performed the analytic integration with respect to $\varphi$  in the quasi-static limit
$$\int\mathbf{E}_{\mathbf{k}}^{\pm}d\varphi=2\pi B_{k}^{\pm}\left\{ \begin{array}{cc}
\pm e^{k\left(z+d\right)}\left\{ \begin{array}{c}
i\cosh\left(kd\right)\\
-\sinh\left(kd\right)
\end{array}\right\} \left(-\mathbf{e}_{z}J_{0}\left(k\rho\right)+\mathbf{e_{\boldsymbol{\rho}}}J_{1}\left(k\rho\right)\right) & \mathbf{r}\in\mathbf{\mathrm{I}}\\
\mathbf{e}_{z}\left\{ \begin{array}{c}
-i\sinh\left(kd\right)\\
\cosh\left(kd\right)
\end{array}\right\} J_{0}\left(k\rho\right)+\mathbf{e_{\boldsymbol{\rho}}}iJ_{1}\left(k\rho\right)\left\{ \begin{array}{c}
\cosh\left(kd\right)\\
i\sinh\left(kd\right)
\end{array}\right\}  & \mathbf{r}\in\mathbf{\mathrm{II}}\\
e^{-k\left(z-d\right)}\left\{ \begin{array}{c}
i\cosh\left(kd\right)\\
-\sinh\left(kd\right)
\end{array}\right\} \left(\mathbf{e}_{z}J_{0}\left(k\rho\right)+\mathbf{e_{\boldsymbol{\rho}}}J_{1}\left(k\rho\right)\right) & \mathbf{r}\in\mathbf{\mathrm{III}}
\end{array}\right..$$

\subsubsection{Region I}
The integrand in Reg. I is 

$$\frac{\frac{s_{\mathbf{k}}}{s-s_{\mathbf{k}}}\left\langle \mathbf{E_{-k}^{\mp}}|\mathbf{E}_{0}\right\rangle \int E_{\mathrm{\mathbf{k},I}}^{\mp}\left(\mathbf{r}\right)kd\varphi}{\left(2\pi\right)^{2}\langle\tilde{{\bf E}}_{{\bf k}}^{\mp}|{\bf E}_{{\bf k}}^{\mp}\rangle}=\frac{k^{2}p\left(e^{4dk}-1\right)e^{k(z-z_{0})}}{\epsilon_{2}\left(4\Delta s^{2}e^{4dk}-1\right)}\left[-J_{0}(k\rho)\hat{z}+J_{1}(k\rho)\hat{\rho}\right].$$
For $\Delta s=0$ we get: 
\begin{gather}
\left.\frac{\frac{s_{\mathbf{k}}}{s-s_{\mathbf{k}}}\left\langle \mathbf{E_{-k}^{\mp}}|\mathbf{E}_{0}\right\rangle \int E_{\mathrm{\mathbf{k},I}}^{\mp}\left(\mathbf{r}\right)kd\varphi}{\left(2\pi\right)^{2}\langle\tilde{{\bf E}}_{{\bf k}}^{\mp}|{\bf E}_{{\bf k}}^{\mp}\rangle}\right|_{\Delta s=0}\nonumber\\
=\frac{k^{2}p\left(e^{4dk}-1\right)J_{0}(k\rho)e^{k(z-z_{0})}}{\epsilon_{2}}\hat{z}-\frac{k^{2}p\left(e^{4dk}-1\right)J_{1}(k\rho)e^{k(z-z_{0})}}{\epsilon_{2}}\hat{\rho}.
\end{gather}

The integral with respect to $\left|\mathbf{k}\right|$  diverges for $z>z_{0}-4d,$ which is above the geometric optics image (see Fig. 1).
 We integrate analytically with respect to $\left|\mathbf{k}\right|$ for $z<z_{0}-4d,$ where the integral converges
\begin{gather}
\int\left(\left.\frac{\frac{s_{\mathbf{k}}}{s-s_{\mathbf{k}}}\left\langle \mathbf{E_{-k}^{\mp}}|\mathbf{E}_{0}\right\rangle \int E_{\mathrm{\mathbf{k},I}}^{\mp}\left(\mathbf{r}\right)kd\varphi}{\left(2\pi\right)^{2}\langle\tilde{{\bf E}}_{{\bf k}}^{\mp}|{\bf E}_{{\bf k}}^{\mp}\rangle}\right|_{\Delta s=0}\right)dk\nonumber \\
=\frac{p}{\epsilon_{2}}\left\{ \left[\frac{3\left(4d+z-z_{0}\right)^{2}}{\left(\left(4d+z-z_{0}\right)^{2}+\rho^{2}\right)^{5/2}}-\frac{1}{\left(\left(4d+z-z_{0}\right)^{2}+\rho^{2}\right)^{3/2}}-\left(\frac{3\left(z-z_{0}\right)^{2}}{\left(\rho^{2}+\left(z-z_{0}\right)^{2}\right)^{5/2}}-\frac{1}{\left(\rho^{2}+\left(z-z_{0}\right)^{2}\right)^{3/2}}\right)\right]\hat{z}\right.\nonumber \\
\left.+\left[\frac{3\rho(4d+z-z_{0})}{\left(\rho^{2}+\left(4d+z-z_{0}\right)^{2}\right)^{5/2}}-\frac{3\rho(z-z_{0})}{\left(\rho^{2}+\left(z-z_{0}\right)^{2}\right)^{5/2}}\right]\hat{\rho}\right\} .\nonumber
\end{gather} 

Adding to this expression 
$$\mathbf{E}_{0}=\frac{p}{\epsilon_{2}}\frac{1}{\left(\rho^{2}+\left(z-z_{0}\right)^{2}\right)^{3/2}}\left\{ \left[\frac{3\left(z-z_{0}\right)^{2}}{\rho^{2}+\left(z-z_{0}\right)^{2}}-1\right]\hat{z}+3\frac{\rho\left(z-z_{0}\right)}{\rho^{2}+\left(z-z_{0}\right)^{2}}\hat{\rho}\right\} ,$$
 we obtain
$$\left.\mathbf{E}_{\mathrm{\mathrm{I}}}\right|_{\Delta s=0}=\frac{p}{\epsilon_{2}}\left\{ \left[\frac{3\left[z-\left(z_{0}-4d\right)\right]^{2}}{\left(\left[z-\left(z_{0}-4d\right)\right]^{2}+\rho^{2}\right)^{5/2}}-\frac{1}{\left(\left[z-\left(z_{0}-4d\right)\right]^{2}+\rho^{2}\right)^{3/2}}\right]\hat{z}+\left[\frac{3\rho\left[z-\left(z_{0}-4d\right)\right]}{\left(\rho^{2}+\left[z-\left(z_{0}-4d\right)\right]{}^{2}\right)^{5/2}}\right]\hat{\rho}\right\},$$
 which is the electric field of an electric point dipole located at $z=z_{0}-4d,$  oriented along the $z$ axis.
\subsubsection{Region II}
The integrand in Reg. II is
\begin{gather}
\frac{\frac{s_{\mathbf{k}}}{s-s_{\mathbf{k}}}\left\langle \mathbf{E_{-k}^{\mp}}|\mathbf{E}_{0}\right\rangle \int E_{\mathrm{\mathbf{k},II}}^{\mp}\left(\mathbf{r}\right)kd\varphi}{\left(2\pi\right)^{2}\langle\tilde{{\bf E}}_{{\bf k}}^{\mp}|{\bf E}_{{\bf k}}^{\mp}\rangle}\nonumber\\
=\frac{k^{2}pe^{k(d-z_{0})}\cosh(dk)\text{csch}(2dk)J_{0}(k\rho)\left[\frac{\left(e^{2dk}-1\right)^{2}\sinh(kz)}{2\Delta se^{4dk}+e^{2dk}}+\frac{2\sinh(2dk)\cosh(kz)}{2\Delta se^{2dk}-1}\right]}{\epsilon_{2}}\hat{z}\nonumber\\
-\frac{k^{2}pJ_{1}(k\rho)\left\{ e^{2dk}\left[2\Delta s\left(e^{2k(d+z)}-1\right)-1\right]+e^{2kz}\right\} e^{-k(z+z_{0})}}{\text{\ensuremath{\epsilon_{2}}}\left(4\Delta s^{2}e^{4dk}-1\right)}\hat{\rho}.
\end{gather}

For $\Delta s=0$ we obtain
\begin{gather}
\left.\frac{\frac{s_{\mathbf{k}}}{s-s_{\mathbf{k}}}\left\langle \mathbf{E_{-k}^{\mp}}|\mathbf{E}_{0}\right\rangle \int E_{\mathrm{\mathbf{k},II}}^{\mp}\left(\mathbf{r}\right)kd\varphi}{\left(2\pi\right)^{2}\langle\tilde{{\bf E}}_{{\bf k}}^{\mp}|{\bf E}_{{\bf k}}^{\mp}\rangle}\right|_{\Delta s=0}\nonumber\\
=-\frac{p}{\epsilon_{2}}\left\{ k^{2}J_{0}(k\rho)]\left[e^{k(2d-z-z_{0})}+e^{k(z-z_{0})}\right]\hat{z}+k^{2}J_{1}(k\rho)\left[e^{k(2d-z-z_{0})}-e^{k(z-z_{0})}\right]\hat{\rho}\right\} .
 \end{gather}

We add to this expression $\mathbf{E}_{0}$ and integrate analytically with respect to $\left|\mathbf{k}\right|.$
This integral diverges for $z<2d-z_{0},$ which is below the geometric optics image.
 For $z>2d-z_{0},$ where the integral converges, we obtain 
$$\left.\mathbf{E}_{\mathrm{\mathrm{II}}}\right|_{\Delta s=0}=-\frac{p}{\epsilon_{2}}\left\{ \left[\frac{\left[3z-\left(2d-z_{0}\right)\right]^{2}}{\left(\left[z-\left(2d-z_{0}\right)\right]^{2}+\rho^{2}\right)^{5/2}}-\frac{1}{\left(\left[z-\left(2d-z_{0}\right)\right]^{2}+\rho^{2}\right)^{3/2}}\right]\hat{z}+\left[\frac{3\rho\left[z-\left(2d-z_{0}\right)\right]}{\left(\left[z-\left(2d-z_{0}\right)\right]^{2}+\rho^{2}\right)^{5/2}}\right]\hat{\rho}\right\} ,$$
which is the electric field of an electric point dipole located at $z=2d-z_{0},$
 directed in $-\hat{z}$ direction.

\subsubsection{Region III}
The integrand in Reg. III is
\[
\frac{\frac{s_{\mathbf{k}}}{s-s_{\mathbf{k}}}\left\langle \mathbf{E_{-k}^{\mp}}|\mathbf{E}_{0}\right\rangle \int E_{\mathrm{\mathbf{k},III}}^{\mp}\left(\mathbf{r}\right)kd\varphi}{\left(2\pi\right)^{2}\langle\tilde{{\bf E}}_{{\bf k}}^{\mp}|{\bf E}_{{\bf k}}^{\mp}\rangle}=-\frac{4p\Delta s\sinh\left(2kd\right)e^{4dk}k^{2}e^{-k(z+z_{0})}}{\epsilon_{2}\left(4\Delta s^{2}e^{4dk}-1\right)}\left[J_{0}(k\rho)\hat{z}+J_{1}(k\rho)\hat{\rho}\right]
\]
For $\Delta s=0$ we get
\[
\left.\frac{\frac{s_{\mathbf{k}}}{s-s_{\mathbf{k}}}\left\langle \mathbf{E_{-k}^{\mp}}|\mathbf{E}_{0}\right\rangle \int E_{\mathrm{\mathbf{k},III}}^{\mp}\left(\mathbf{r}\right)kd\varphi}{\left(2\pi\right)^{2}\langle\tilde{{\bf E}}_{{\bf k}}^{\mp}|{\bf E}_{{\bf k}}^{\mp}\rangle}\right|_{\Delta s=0}=0.
\]
Therefore, the electric field for $\Delta s=0$ is
\[
\left.\mathbf{E}_{\mathrm{III}}\right|_{\Delta s=0}=\mathbf{E}_{0}=\frac{p}{\epsilon_{2}}\frac{1}{\left(\rho^{2}+\left(z-z_{0}\right)^{2}\right)^{3/2}}\left\{ \left[\frac{3\left(z-z_{0}\right)^{2}}{\left(\rho^{2}+\left(z-z_{0}\right)^{2}\right)}-1\right]\hat{z}+3\frac{\rho\left(z-z_{0}\right)}{\left(\rho^{2}+\left(z-z_{0}\right)^{2}\right)}\hat{\rho}\right\} ,
\]
\end{widetext}
which is the electric field of the electric point dipole located at $z=z_{0},$ oriented along the $z$ axis.

 For $\Delta s=0$ the regions where the electric field diverges are between the images expected according to geometric optics in Regions I and II. This is in agreement with the conclusions in Ref. \cite{BergPRA2014} where a point charge object was considered.  For $\Delta s=0$ the electric field (where it does not diverge) in Reg I,II and III is equal to the electric field of point dipoles located at the geometric image foci directed in the $\hat{z},-\hat{z}$ and $\hat{z}$ directions respectively. This is in agreement with the results in Ref. \cite{BergPRA2014} in which the electric field of a point charge object and $\Delta s=0,$ in Regions I,II and III is equal to the electric field of point charges located at the geometric image foci.




\end{document}